\documentclass[10pt,a4paper,logo]{paper}

\usepackage[all]{hypcap}

\usepackage[authoryear, round]{natbib}

\usepackage{nicefrac}%
\usepackage{multirow}
\usepackage{cleveref}
\usepackage{dsfont}
\usepackage{makecell}
\usepackage{subcaption}
\usepackage{bbm}
\usepackage{wrapfig}
\usepackage{mathtools,xparse}
\usepackage{array}
\usepackage{scalerel}
\usepackage{tablefootnote}
\usepackage{tocloft}
\usepackage{adjustbox}
\usepackage{setspace}
\usepackage[most,skins,theorems]{tcolorbox}

\usepackage{float}
\usepackage{afterpage}
\usepackage{tabularx}
\newcolumntype{Y}{>{\centering\arraybackslash}X}%

\definecolor{tolblue}{HTML}{4477AA}
\definecolor{tolcyan}{HTML}{66CCEE}
\definecolor{tolgreen}{HTML}{228833}
\definecolor{tolyellow}{HTML}{CCBB44}
\definecolor{tolred}{HTML}{EE6677}
\definecolor{tolpurple}{HTML}{AA3377}
\definecolor{tolgrey}{HTML}{BBBBBB}

\usepackage{fvextra}
\DefineVerbatimEnvironment{promptbox}{Verbatim}{
  fontsize={\fontsize{8pt}{9.5pt}\selectfont},
  breaklines=true,
  breaksymbolleft={},
  breakindent=0pt,
  frame=single,
  framesep=3pt
}
\DeclareUnicodeCharacter{2113}{\ensuremath{\ell}}
\DeclareUnicodeCharacter{2010}{-}

\newboolean{showsection}
\setboolean{showsection}{true}
\makeatletter
\@namedef{ver@everyshi.sty}{}
\makeatother

\definecolor{custom_green}{rgb}{0.0, 0.5, 0.0}
\definecolor{custom_red}{rgb}{1.0, 0.01, 0.24}
\definecolor{custom_blue}{HTML}{C9DAF7}
\definecolor{custom_purple}{HTML}{D9D1E9}
\definecolor{title_blue}{HTML}{204899}
\definecolor{cite_blue}{HTML}{044dc1}
\definecolor{cite_purple}{HTML}{7406a7}

\hypersetup{
    colorlinks = true,
    citecolor = {cite_blue},
    linkcolor = {cite_purple},
    urlcolor = {cite_purple},
}

\definecolor{blanchedalmond}{rgb}{1.0, 0.92, 0.8}
\definecolor{carmine}{rgb}{0.59, 0.0, 0.09}
\definecolor{lightblue}{rgb}{0.22,0.45,0.70}%

\renewcommand{\mathbf}{\boldsymbol}

\makeatletter
\def\Ddots{\mathinner{\mkern1mu\raise\p@
\vbox{\kern7\p@\hbox{.}}\mkern2mu
\raise4\p@\hbox{.}\mkern2mu\raise7\p@\hbox{.}\mkern1mu}}
\makeatother

\numberwithin{equation}{section}

\definecolor{amaranth}{rgb}{0.9, 0.17, 0.31}
\definecolor{antiquebrass}{rgb}{0.8, 0.58, 0.46}
\definecolor{antiquefuchsia}{rgb}{0.57, 0.36, 0.51}
\definecolor{chromeyellow}{rgb}{0.31, 0.47, 0.26}

\newcommand{\1}{\mathds 1}

\usepackage{amsmath,amsfonts,bm}

\def\eqref#1{equation~\ref{#1}}

\def\1{\bm{1}}

\DeclareMathAlphabet{\mathsfit}{\encodingdefault}{\sfdefault}{m}{sl}
\SetMathAlphabet{\mathsfit}{bold}{\encodingdefault}{\sfdefault}{bx}{n}

\makeatletter
\def\mathcolor#1#{\@mathcolor{#1}}
\def\@mathcolor#1#2#3{%
  \protect\leavevmode
  \begingroup
    \color#1{#2}#3%
  \endgroup
}
\makeatother

\Crefformat{equation}{#2Eq.\;(#1)#3}
\Crefformat{figure}{#2Figure #1#3}
\Crefformat{assumption}{#2Assumption #1#3}
\Crefname{assumption}{Assumption}{Assumptions}

\usepackage{crossreftools}
\makeatletter
\renewcommand\footnoterule{%
  \kern 15\p@
  \hrule \@width 2in \kern 2.6\p@
  \vspace{4pt}
}
\makeatother

\title{Cross-Pollination Retrieval: Finding Transferable Ideas Across Fields in Scientific Literature}

\author[1]{Seokwon~Jung}
\author[1]{Yerang~Kim}
\author[1]{Yejin~Kim}
\author[2]{Alexander~Rubinstein}
\author[2,\textbf{*}]{Arnas~Uselis}
\author[1,\textbf{*}]{Sangwoo~Park}
\author[1]{Seong~Joon~Oh}
\author[3]{Sangdoo~Yun}

\affil[1]{KAIST}
\affil[2]{T\"ubingen AI Center, University of T\"ubingen}
\affil[3]{NAVER AI Lab}

\makeatletter\@ifundefined{correspondingauthor}{}{\correspondingauthor{Correspondence to: \href{mailto:SeokwonJung@kaist.ac.kr}{\texttt{SeokwonJung@kaist.ac.kr}}.}}\makeatother

\begin{abstract}
Many scientific advances come from cross-pollination, carrying an idea from one field over to a problem in another. Yet retrieving papers across fields remains underexplored, with existing benchmarks confined to a narrow range of fields. We study cross-pollination retrieval, the problem of identifying transferable ideas across research fields, and introduce CPR-Dataset, a large-scale dataset constructed by mining cross-topic citations. Its evaluation split CPR-eval has 1,300 queries in 26 fields, and its training split CPR-train has 10,573 pairs. On CPR-eval, the best of eleven existing retrievers reaches a Recall@10 of only 0.26, whereas an LLM reranker reaches 0.81. Standard contrastive fine-tuning on CPR-train raises Recall@10 to as high as 0.70. The gain carries over to domains held out from training and to analogy and reasoning-intensive retrieval. On AnaloBench, it raises nDCG@10 by up to 33\% over the untrained retriever, whereas training on three existing retrieval datasets lowers it. On BRIGHT, it raises nDCG@10 by up to 70\%, more than twice the gain from those datasets. CPR-Dataset is available at \href{https://huggingface.co/CPR-Dataset}{\texttt{huggingface.co/CPR-Dataset}}, and the code at \href{https://github.com/SeokwonJung-Jay/Cross-Pollination-Retrieval}{\texttt{github.com/SeokwonJung-Jay/Cross-Pollination-Retrieval}}.
\end{abstract}

\begin{document}

\maketitle

\begin{figure}[!b]
\centering
\includegraphics[width=\linewidth]{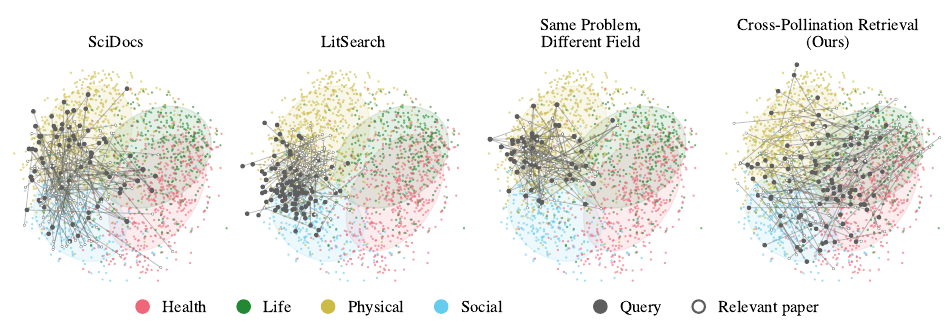}
\caption{\textbf{Queries and their relevant papers stay in a narrow region of science in existing benchmarks, whereas those of our Cross-Pollination Retrieval benchmark span all four domains.} Each panel shows the 100 pairs farthest apart on a PCA map of paper embeddings (E5-base-v2~\citep{wang2024textembeddingsweaklysupervisedcontrastive}) in SciDocs~\citep{cohan-etal-2020-specter}, LitSearch~\citep{ajith-etal-2024-litsearch}, Same Problem, Different Field~\citep{kulikowski2026problemdifferentfieldcrossdomain}, or ours, with background papers and ellipses colored by OpenAlex domain. Lines join each query (filled dot) to its relevant paper (open circle).}
\label{fig:landscape}
\end{figure}
\section{Introduction}
\label{sec:intro}

\begin{figure}[t]
\centering
\includegraphics{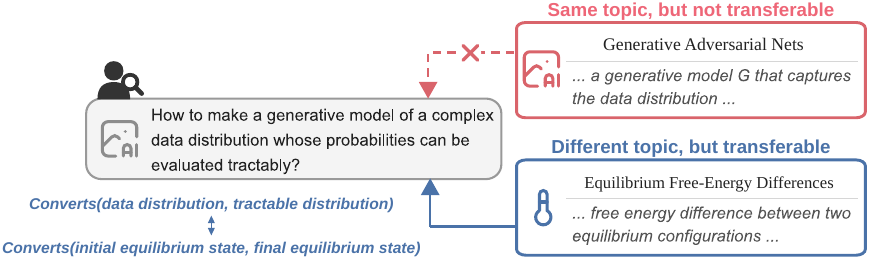}
\caption{\textbf{Example of a cross-pollination pair.}
The query describes the problem of \citet{sohldickstein2015deep}, which introduced diffusion models.
A same-topic paper, generative adversarial nets~\citep{goodfellow2014gan}, cannot be transferred to it.
A paper from nonequilibrium statistical physics~\citep{jarzynski1997equilibrium} can: it moves a system from an initial to a final equilibrium state, as the base paper turns the data distribution into a tractable one.
The query and titles are shortened.}
\label{fig:example}
\end{figure}

Many impactful scientific discoveries come from \textit{cross-pollination}: transferring a research idea from one field to another.
An analysis of 9.2 million papers across 13 fields found that papers co-citing the least related areas of research received 72\% more citations than average~\citep{lariviere2015longdistance}.
Similarly, an analysis of 17.9 million papers found that papers combining conventional prior work with a few rarely combined areas of research were nearly twice as likely to be highly cited~\citep{uzzi2013atypical}.
However, as knowledge accumulates, researchers specialize in increasingly narrow areas of expertise~\citep{jones2009burden}.
Retrieval systems that surface cross-field connections could thus expose researchers to useful solutions from other fields.

\begin{wrapfigure}{r}{0.48\linewidth}
\centering
\includegraphics[width=\linewidth]{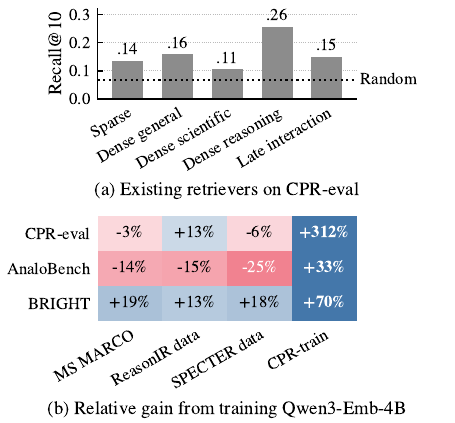}
\caption{\textbf{Existing retrievers fail at cross-pollination retrieval, and training on CPR-train helps beyond CPR-eval.} (a) Recall@10 on CPR-eval, averaged within each retriever family; the dotted line marks random ranking. \textit{Dense scientific} and \textit{Dense reasoning} denote retrievers trained for scientific literature search and for reasoning-intensive retrieval. (b) Relative gain of Qwen3-Embedding-4B over its untrained version after one epoch on the same number of pairs from each dataset (Recall@10 on CPR-eval, nDCG@10 on AnaloBench and BRIGHT); for existing datasets, the mean over three random samples. Colors are scaled within each row.}
\label{fig:motivation}
\end{wrapfigure}
Yet retrieving such connections across fields remains underexplored. Current automated approaches rely on retrievers to find work that may be applicable to a given problem~\citep{liu-etal-2026-researchbench,yang2025moosechem,kim2026scholarcatalystbenchmarkretrievingpapers,pandey2026musesbenchmarkprospectiveintellectualroots,asai2024openscholarsynthesizingscientificliterature}. However, the benchmarks used to evaluate these retrievers pair queries with relevant papers that are semantically similar and limited in field coverage (Figure~\ref{fig:landscape}). In this work, we extend the field coverage and ask whether current retrievers can support such \textit{cross-pollination retrieval}: finding papers across fields whose ideas can be transferred. 
To study this question, we build a Cross-Pollination Retrieval benchmark (CPR-eval, Figure~\ref{fig:example}) with a pipeline that mines citations through which a paper transferred an idea from another topic.
The pipeline scales CPR-eval to 1,300 queries across 26 fields, 14$\times$ more queries and 1.9$\times$ more fields than the closest prior work (Appendix~\ref{app:comparison}).

Evaluating existing retrievers on CPR-eval, we find that they fail at cross-pollination retrieval (Figure~\ref{fig:motivation}a).
On average, existing retrievers reach only about twice the Recall@10 of random ranking.
Surprisingly, retrievers built specifically for scientific literature retrieval perform worse on average than general-purpose ones.

We hypothesize that retrievers fail on cross-pollination retrieval because they lack training data for it.
We therefore construct CPR-train, a large training corpus of 10,573 pairs, each linking a research problem to a paper on another topic whose idea was transferred to it.
Fine-tuning three retrievers, Qwen3-Embedding-0.6B and 4B and llama-embed-nemotron-8b, on CPR-train substantially improves all three on CPR-eval, whereas training them on existing retrieval datasets brings little gain (Figure~\ref{fig:motivation}b).
CPR-train also generalizes beyond its own distribution.
Qwen3-Embedding-0.6B and 4B trained without one of the four domains, the top level of the OpenAlex taxonomy~\citep{openalex2024topics} (health, life, physical, and social sciences), still improve on the held-out domain.
Importantly, the trained retrievers improve on AnaloBench~\citep{ye-etal-2024-analobench} and BRIGHT~\citep{su2025bright} more than the same retrievers trained on existing data: web search queries (MS MARCO~\citep{DBLP:journals/corr/NguyenRSGTMD16}), scientific citations (SPECTER data~\citep{cohan-etal-2020-specter}), or data built for reasoning-intensive retrieval (ReasonIR data~\citep{shao2025reasonir}). We note that these gains come from standard contrastive fine-tuning without architectural modifications, which highlights the value of CPR-train as training data.

In summary, our contributions are as follows:
\begin{itemize}
  \item We introduce cross-pollination retrieval and release CPR-Dataset, built scalably from citations: CPR-eval with 1,300 queries across 26 fields, and CPR-train with 10,573 pairs.
  
  \item We show that existing retrievers fail at cross-pollination retrieval, and retrievers built for scientific literature perform worst on average.
  
  \item We show that training on CPR-train improves retrievers more than training on existing data, on CPR-eval, held-out domains, AnaloBench, and BRIGHT. This establishes CPR-train as an effective resource for training retrievers beyond topical similarity.

\end{itemize}
We release CPR-Dataset and the pipeline that collects it. We will release the Qwen3-Embedding retrievers trained on CPR-train upon publication, and hope that these resources drive new research on cross-pollination retrieval.

\section{Related Work}
\label{sec:related}
\paragraph{Inspiration retrieval for scientific discovery.}
In scientific literature retrieval, inspiration retrieval is the task of finding prior papers that can inspire new research ideas.
ScholarCatalyst~\citep{kim2026scholarcatalystbenchmarkretrievingpapers} defines an inspiration as a prior paper that, according to the authors of a project, helped or could have helped advance it.
MUSES~\citep{pandey2026musesbenchmarkprospectiveintellectualroots} defines it as a prior paper that the authors cite in their next paper.
MOOSE-Chem~\citep{yang2025moosechem} and ResearchBench~\citep{liu-etal-2026-researchbench} define it as a reference that domain experts or an LLM, respectively, judge necessary to form the main hypothesis of a paper.
None of these definitions requires an inspiration to come from a field other than that of the query.
Some work does look for papers in other fields: SOLVENT~\citep{10.1145/3274300} finds analogies between papers across domains, and Same Problem, Different Field~\citep{kulikowski2026problemdifferentfieldcrossdomain} finds papers in other fields that solve the same computational problem.
Because finding such matches by hand is costly, both are evaluated on small curated sets of at most 109 papers.
In contrast, CPR-Dataset requires every positive to come from another topic and builds such pairs at scale from citations.

\paragraph{Scientific literature retrievers.}
Most retrievers for scientific literature learn relevance from citations.
SPECTER~\citep{cohan-etal-2020-specter} treats a paper that the query paper cites as a positive and an uncited paper as a negative, SciNCL~\citep{ostendorff-etal-2022-neighborhood} samples positives and negatives from neighborhoods in a citation embedding space, and SPECTER2~\citep{singh-etal-2023-scirepeval} extends this training to multiple task formats.
The OpenScholar retriever~\citep{asai2024openscholarsynthesizingscientificliterature} instead adapts to scientific text without relevance labels, continuing the unsupervised contrastive pre-training of Contriever~\citep{izacard2022unsupervised} on scientific papers.
None of these retrievers is designed for cross-pollination retrieval, for which we build CPR-eval as an evaluation set and CPR-train as a training set.

\paragraph{Retrieval beyond semantic similarity.}
Several lines of work define relevance beyond the semantic similarity of content.
Reasoning-intensive retrieval benchmarks~\citep{su2025bright,zhao-etal-2026-rethinking} are designed so that the relevant documents for a query can be identified only through additional reasoning steps.
Specifically, BRIGHT pairs natural-language questions from seven StackExchange sites with the web pages cited in their answers, alongside coding and mathematics subsets that rely on domain-specific notation~\citep{zhao-etal-2026-rethinking}.
Analogy datasets~\citep{ye-etal-2024-analobench,jiayang-etal-2023-storyanalogy} instead pair texts that share the same underlying relational structure.
In AnaloBench, the authors write pairs of analogous sentences on different topics, and GPT-4 expands each sentence into stories of 10 and 30 sentences.
We define cross-pollination retrieval as another line, in which relevant papers come from a different topic yet offer insights transferable to the research problem of the query.
We also test whether what retrievers learn from CPR-train extends to the other two lines.

\section{Cross-Pollination Retrieval}
\label{sec:dataset}

\begin{figure}[t]
\centering
\includegraphics{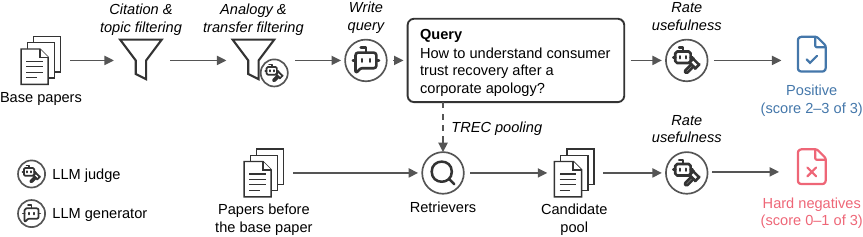}
\caption{\textbf{Construction pipeline of CPR-Dataset.}
Top: cited papers on other topics are filtered for analogy and transfer, and those rated useful for the generated query become positives (Section~\ref{sec:positives}).
Bottom: hard negatives are candidates pooled by three retrievers and rated not useful (Section~\ref{sec:negatives}).}
\label{fig:pipeline}
\end{figure} 

We build CPR-Dataset in three stages (Figure~\ref{fig:pipeline}): we define relevance through analogy to go beyond semantic similarity (Section~\ref{sec:task}), mine positives from citations to ground them in transfers that authors actually made (Section~\ref{sec:positives}), and collect same-topic hard negatives so that topic alone does not suffice (Section~\ref{sec:negatives}).

\noindent

\subsection{Definition of the Cross-Pollination Retrieval Task}
\label{sec:task}

Given a research problem, cross-pollination retrieval is the task of finding papers on other topics whose ideas can be transferred to that problem. We denote as $q$ a base paper addressing a research problem $p_q$, and let $x_q$ be a natural-language query describing this problem. We aim to retrieve a paper $d$ addressing another research problem $p_d$, whose ideas can help solve $p_q$. 

Finding transferable ideas across fields means that:
(1) The retrieved paper has to come from outside the original problem's field. We argue that a good proxy for quantifying this is requiring that no topics describing the two papers are shared. (2) However, being from different topics does not tell us how an
idea developed for one problem could be applied to another. We focus on cases where the two problems share some underlying structure, allowing us to identify correspondences between them and explain how a solution developed for one could be transferred
to the other. We characterize such connections  through structural analogy. 
(3) Finally, a correspondence does not guarantee that the transfer is useful. Even if two problems are analogous, the solution developed for one might not help solve the other. We therefore also require that the retrieved paper's contribution, carried through this correspondence, helps solve the original problem. We formalize these conditions below.

\paragraph{1) Cross-topic.}
$d$ shares no assigned topic with $q$, $T(q) \cap T(d) = \emptyset$, where $T(\cdot)$ denotes the set of topics that a topic classifier assigns to a paper (e.g., \textit{Gaussian Processes and Bayesian Inference} for the base paper in Figure~\ref{fig:example} and \textit{Advanced Thermodynamics and Statistical Mechanics} for the relevant paper; more examples in Table~\ref{tab:openalex-examples}).

\paragraph{2) Analogous.}

Following \citet{shen2026unlockingllmcreativityscience} and structure-mapping theory~\citep{gentner1983structure}, we represent the research problem of $q$ as $p_q = (O_q, R_q)$, and likewise $p_d = (O_d, R_d)$ for $d$.
$O_q$ is the set of objects the problem involves and $R_q$ is the set of relations among them, each of the form $r(o_1, \dots, o_k)$ with a relation type $r$, $k \geq 2$, and $o_i \in O_q$.

An analogy $\mathcal{A} = (M, S)$ consists of an object mapping $M : O_q \rightarrow O_d$, which gives each object of $p_q$ at most one counterpart in $p_d$, and the set of relations of $p_q$ that also hold in $p_d$ under $M$:
\[
S = \{\, r(\mathbf{o}) \in R_q : r(M(\mathbf{o})) \in R_d \,\},
\]
where $M(\mathbf{o})$ applies $M$ to each argument of $\mathbf{o} = (o_1, \dots, o_k)$.
$d$ is analogous to $q$ if there is an analogy with $|M| \geq 2$ and $S \neq \emptyset$.

For example, the base paper in Figure~\ref{fig:example} learns a generative model by gradually converting a complex data distribution into a tractable one, and the relevant paper drives a physical system from one equilibrium state to another.
The bottom left of the figure shows their analogy: the same relation, \textsc{Converts}, holds between the two distributions and between the two equilibrium states. Formally,
\begin{align*}
M &= \{\text{data dist.} \mapsto \text{initial equilibrium state},\ \text{tractable dist.} \mapsto \text{final equilibrium state}\},\\
S &= \{\textsc{Converts}(\text{data dist.}, \text{tractable dist.})\}.
\end{align*}
See Appendix~\ref{app:analogy-example} for more examples.

\paragraph{3) Useful.}
$d$ is useful to $q$ if its contribution, a method, finding, or formulation, carried through $\mathcal{A}$, can be used to solve $p_q$.

\subsection{Positive Collection}
\label{sec:positives}

We collect positive pairs in the following three steps:

\paragraph{Step P1: Cross-topic citation mining.} We take the topics of each paper from \citet{openalex2024topics}, which organizes research into 4 domains, 26 fields, 252 subfields, and about 4,500 topics. It assigns up to three topics to each paper using a classifier that combines a fine-tuned multilingual BERT encoder of its title and abstract with its journal name and citations.
We collect the papers for dataset construction from Semantic Scholar~\citep{kinney2023semantic}, which provides both their full text and citation information.
For each of the 26 OpenAlex fields, we take the venues in the top 20\% by impact factor and sample the same number of base papers from their 2010--2025 publications.
Among the papers $d$ cited by each base paper $q$, we extract the pairs whose topics do not overlap.

\paragraph{Step P2: Analogy judgment.}
An LLM reads the titles and abstracts of $q$ and $d$ and the sentences in $q$ that cite $d$.
It extracts the object mapping $M$ and shared relations $S$ of Section~\ref{sec:task} and keeps the pair only if they form an analogy and $q$ uses, or could use, the contribution of $d$ through it (Appendix~\ref{app:prompt-analogy}).
This judgment discards about 91\% of the cross-topic citation pairs, mostly because $q$ cites $d$ only as background.
A second LLM call checks that $q$ actually uses the contribution of $d$, rejecting pairs that address the same problem, cite $d$ only as background, or rest on unsupported mappings (Appendix~\ref{app:prompt-transfer}). This verification agrees with a human annotator at Cohen's $\kappa$ = 0.84 (Appendix~\ref{app:human-agreement}).

\paragraph{Step P3: Query generation and verification.}
For each remaining pair, an LLM writes a query $x_q$ that states only the problem of $q$ (Appendix~\ref{app:prompt-query}). It reads the abstract of $q$ and the sentences citing $d$, with the citation masked, and leaves out the contribution of $d$. When several queries from the same $q$ ask about the same problem, an LLM keeps the pair with the strongest analogy, carrying over most of $d$'s method, formulation, or finding (Appendix~\ref{app:prompt-dedup}). Finally, since the query omits the citation context, an LLM rates on a 0--3 scale how well $d$ can solve the problem stated in $x_q$, giving $u(q,d)$, and we keep only pairs with $u(q,d) \geq 2$ (Appendix~\ref{app:prompt-usefulness}). Binarized at 2, these scores agree with a human annotator at Cohen's $\kappa$ = 0.84 (Appendix~\ref{app:human-agreement}).

\subsection{Hard Negative Collection}
\label{sec:negatives}

Hard negatives are papers that look relevant to the query but cannot be used to solve the problem of $q$.
They serve two purposes.
In evaluation, they keep a retriever from succeeding through topical similarity alone; benchmarks for reasoning-intensive retrieval and inspiration retrieval likewise add candidates that are related to the query but do not meet its need~\citep{kim2026scholarcatalystbenchmarkretrievingpapers,su2025bright,liu-etal-2026-researchbench}.
In training, they keep a retriever from learning the shortcut of ranking by surface similarity: retrievers for reasoning-intensive retrieval are likewise trained with negatives that look relevant to the query but do not help answer it~\citep{shao2025reasonir,DIVER}.
To collect such negatives, we first pool candidates that look relevant to the query (Step N1), and then we judge every candidate before using it as a negative (Step N2).

\paragraph{Step N1: Candidate pooling.}
We draw candidates from the papers in the top 20\% venues of Section~\ref{sec:positives} and the papers they cite, keeping those with an abstract.
For each query, we consider only papers published before $q$, since these were the papers available when its authors faced the problem.
We then split them into three groups by topic: papers sharing a topic with $q$, papers sharing a topic only with the positive $d^{+}$, and the rest.
Each group first takes the papers that $q$ cites but that the analogy judgment of Section~\ref{sec:positives} rejected.
The rest of the group is filled by TREC-style pooling~\citep{voorhees2005trec}: we merge the top 300 papers retrieved by E5~\citep{wang2024textembeddingsweaklysupervisedcontrastive}, SPLADE-v3~\citep{lassance2024spladev3newbaselinessplade}, and SciNCL~\citep{ostendorff-etal-2022-neighborhood} in round-robin order (the top 1,000 for 19 queries whose groups were not filled otherwise).

\paragraph{Step N2: Negative judgment.}
Each candidate is rated with $u(q,d)$ on the same 0--3 scale as the positives (Step P3), but without an analogy, since a candidate has none (Appendix~\ref{app:prompt-hardneg}).
Only papers with $u(q,d) \leq 1$ become hard negatives: going down each group in this order, we keep the first 50 such papers per group in CPR-eval (150 per query) and the first 3 per group in CPR-train (9 per pair). Binarized at 2, these scores agree with a human annotator at Cohen's $\kappa$ = 0.92 (Appendix~\ref{app:human-agreement}).

\subsection{Dataset Statistics}
\label{sec:stats}

\begin{wraptable}{r}{0.48\linewidth}
\vspace{-1.2\baselineskip}
\centering
\small
\begin{tabular*}{\linewidth}{@{\extracolsep{\fill}}lrrrr}
\toprule
Split & Queries & Base papers & Pos. & Neg. \\
\midrule
CPR-eval & 1,300 & 1,080 & 1 & 150 \\
CPR-train & 10,573 & 8,137 & 1 & 9 \\
\bottomrule
\end{tabular*}
\caption{\textbf{Split sizes of CPR-Dataset.} Number of queries and of the distinct base papers they come from, with positives (Pos.) and hard negatives (Neg.) per query.}
\label{tab:splits}
\end{wraptable}
Table~\ref{tab:splits} summarizes CPR-Dataset. CPR-eval has 50 queries from each of the 26 fields, and CPR-train takes the remaining pairs. No base paper or positive of CPR-eval appears in CPR-train, even as a negative. Queries average 20.4 words. Positives lie far from their base papers: 79.9\% come from a different OpenAlex field, and 45.3\% from a different OpenAlex domain (Appendix~\ref{sec:appendix-openalex}, Table~\ref{tab:levels}).

\section{Evaluation and Analysis}
\label{sec:eval}

We evaluate eleven retrievers from five families (sparse, general dense, scientific dense, reasoning dense, and late interaction) on CPR-eval and compare them with an LLM reranker (Section~\ref{sec:main}).
We ask whether this difficulty depends on same-topic candidates, on the pooling retrievers, or on the wording of the query (Section~\ref{sec:difficulty}).
Training on CPR-train partly closes the gap, and we ask how far this improvement generalizes (Section~\ref{sec:generalization}).

\begin{table}[t]
\centering
\small
\setlength{\aboverulesep}{0pt}\setlength{\belowrulesep}{0pt}\renewcommand{\arraystretch}{1.2}
\begin{tabularx}{\linewidth}{l|YY|YY|YY|YY|YY}
\toprule
 & \multicolumn{8}{c|}{\textbf{Domain}} & \multicolumn{2}{c}{} \\
\cmidrule{2-9}
 & \multicolumn{2}{c|}{Health} & \multicolumn{2}{c|}{Life} & \multicolumn{2}{c|}{Physical} & \multicolumn{2}{c|}{Social} & \multicolumn{2}{c}{Avg.} \\
Method & R@1 & R@10 & R@1 & R@10 & R@1 & R@10 & R@1 & R@10 & R@1 & R@10 \\
\midrule
\multicolumn{11}{c}{\textit{LLM reranker}$^{\dagger}$} \\
\midrule
MiMo-V2.6-Pro & 0.60 & 0.86 & 0.60 & 0.80 & 0.56 & 0.77 & 0.58 & 0.86 & 0.58 & 0.81 \\
\midrule
\multicolumn{11}{c}{\textit{Sparse}} \\
\midrule
BM25 & 0.03 & 0.16 & 0.04 & 0.16 & 0.04 & 0.16 & 0.07 & 0.19 & 0.04 & 0.16 \\
BGE-M3 (sparse) & 0.04 & 0.11 & 0.01 & 0.10 & 0.03 & 0.11 & 0.05 & 0.11 & 0.03 & 0.11 \\
\midrule
\multicolumn{11}{c}{\textit{Dense (general)}} \\
\midrule
Qwen3-Emb-0.6B & 0.05 & 0.12 & 0.04 & 0.10 & 0.03 & 0.12 & 0.04 & 0.13 & 0.04 & 0.12 \\
BGE-M3 (dense) & 0.04 & 0.11 & 0.03 & 0.08 & 0.03 & 0.09 & 0.05 & 0.11 & 0.03 & 0.10 \\
Qwen3-Emb-4B & 0.06 & 0.18 & 0.07 & 0.17 & 0.04 & 0.16 & 0.07 & 0.18 & 0.06 & 0.17 \\
Gemini Embedding & 0.10 & 0.25 & 0.12 & 0.26 & 0.09 & \textbf{0.24} & \textbf{0.11} & \textbf{0.29} & 0.10 & \textbf{0.26} \\
\midrule
\multicolumn{11}{c}{\textit{Dense (scientific)}} \\
\midrule
SPECTER2 & 0.02 & 0.11 & 0.03 & 0.10 & 0.02 & 0.09 & 0.01 & 0.07 & 0.02 & 0.09 \\
OpenScholar retriever & 0.02 & 0.11 & 0.03 & 0.10 & 0.04 & 0.13 & 0.05 & 0.14 & 0.04 & 0.12 \\
\midrule
\multicolumn{11}{c}{\textit{Dense (reasoning)}} \\
\midrule
inf-retriever-v1-pro & \textbf{0.13} & \textbf{0.28} & \textbf{0.13} & \textbf{0.28} & \textbf{0.11} & 0.23 & \textbf{0.11} & 0.27 & \textbf{0.12} & \textbf{0.26} \\
\midrule
\multicolumn{11}{c}{\textit{Late interaction}} \\
\midrule
LateOn & 0.10 & 0.19 & 0.08 & 0.18 & 0.09 & 0.19 & 0.09 & 0.21 & 0.09 & 0.19 \\
BGE-M3 (multi-vector) & 0.05 & 0.14 & 0.02 & 0.11 & 0.03 & 0.08 & 0.05 & 0.12 & 0.03 & 0.11 \\
\bottomrule
\end{tabularx}
\caption{\textbf{Recall@1 and Recall@10 of retrievers and an LLM reranker on CPR-eval.} Columns show each OpenAlex domain of the base paper and all queries. Bold marks the best retriever per column. Random ranking gives 0.01 and 0.07. $^{\dagger}$For reference: the LLM reranker reorders the same 151 candidates in a tournament.}
\label{tab:main}
\end{table}

\subsection{Experimental Setup}
\label{sec:setup}

\paragraph{Evaluation.}
Each query is ranked against its 151 candidates: the positive and 150 hard negatives (Section~\ref{sec:negatives}).
Following prior benchmarks~\citep{ye-etal-2024-analobench,liu-etal-2026-researchbench}, we rank only these candidates instead of the whole corpus. Every candidate here has been judged, whereas the corpus may contain unjudged relevant papers.
We report Recall@10 as the main metric, and Recall@1, Recall@5, MRR@10, and the input length of each retriever in Appendix~\ref{sec:appendix-metrics}.

\begin{figure}[t]
\centering
\includegraphics{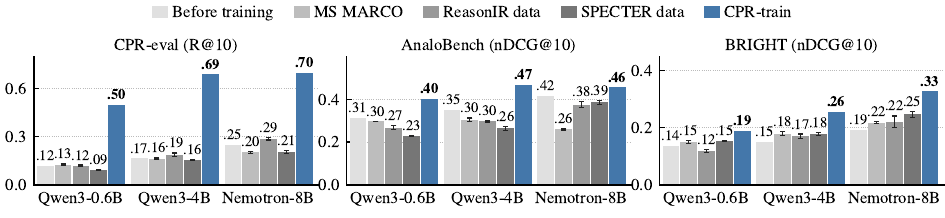}
\caption{\textbf{Retrievers trained on different data.} Each retriever is evaluated before training and after training for one epoch on the same number of pairs (10,573), randomly sampled from each existing dataset; for the existing datasets, bars show the mean over three random samples and whiskers one standard deviation. Qwen3-0.6B and Qwen3-4B denote Qwen3-Embedding-0.6B and 4B, and Nemotron-8B denotes llama-embed-nemotron-8b. BRIGHT covers its StackExchange subsets, and AnaloBench averages stories of 1, 10, and 30 sentences.}
\label{fig:trained}
\end{figure}

\paragraph{Retrievers.}
We evaluate eleven retrievers from five families (Table~\ref{tab:main}).
Sparse retrievers score a paper by the weighted overlap of its terms with the query: BM25~\citep{10.1561/1500000019} and BGE-M3~\citep{chen-etal-2024-m3} in sparse mode.
General dense retrievers encode the query and each paper into a single vector and are not specialized to a domain or task: Qwen3-Embedding-0.6B and 4B~\citep{qwen3embedding}, Gemini Embedding (\texttt{gemini-embedding-001};~\citealp{lee2025geminiembeddinggeneralizableembeddings}), and BGE-M3 in dense mode.
Scientific dense retrievers are trained on scientific papers: SPECTER2~\citep{singh-etal-2023-scirepeval} and the OpenScholar retriever~\citep{asai2024openscholarsynthesizingscientificliterature}.
The reasoning dense retriever, inf-retriever-v1-pro~\citep{infx2025retriever}, is designed for reasoning-intensive queries.
Late-interaction retrievers match the query and the paper token by token: LateOn~\citep{sourty2026denseonlateon} and BGE-M3 in multi-vector mode.
As an LLM reranker, MiMo-V2.6-Pro~\citep{mimo2026v26pro} follows tournament-style reranking~\citep{10.1145/3696410.3714863,OBLIQ-Bench}: it narrows the 151 candidates to ten in two elimination rounds and scores the ten one by one (Appendix~\ref{sec:appendix-reranker}).

\paragraph{Training.}
We fine-tune Qwen3-Embedding-0.6B, Qwen3-Embedding-4B, and llama-embed-nemotron-8b~\citep{babakhin2025llamaembednemotron8buniversaltextembedding} on CPR-train (Section~\ref{sec:stats}) with the InfoNCE loss~\citep{oord2019representationlearningcontrastivepredictive} over hard and in-batch negatives~\citep{karpukhin-etal-2020-dense} (Appendix~\ref{sec:appendix-definitions}).
For comparison, we train the same models on the same number of pairs (10,573), randomly sampled from each of three existing training sets; we draw three samples per set and report the mean (Appendix~\ref{sec:appendix-resample}).
MS MARCO~\citep{DBLP:journals/corr/NguyenRSGTMD16} pairs web search queries with passages that answer them, ReasonIR data~\citep{shao2025reasonir} pairs documents with LLM-generated queries and unhelpful hard negatives, and SPECTER data~\citep{cohan-etal-2020-specter,singh-etal-2023-scirepeval} pairs a paper with a paper it cites and one it does not.
We train every model for one epoch and evaluate its final checkpoint. Hyperparameters are in Appendix~\ref{sec:appendix-hparams}.

\subsection{Main Results}
\label{sec:main}

\paragraph{The task is solvable, but existing retrievers struggle.}
The LLM reranker places the positive in its top 10 for 0.81 of the queries.
The positives are thus recognizable once among the candidates.
Retrievers find far fewer of them (Table~\ref{tab:main}).
The best, Gemini Embedding and inf-retriever-v1-pro, reach 0.26 against 0.07 by chance.
Retrievers trained on scientific papers do no better, with SPECTER2 at 0.09.
The gap holds in every domain.

\paragraph{Training on CPR-train substantially improves retrievers.}
Training on CPR-train raises the Recall@10 of Qwen3-Embedding-0.6B from 0.12 to 0.50 (Figure~\ref{fig:trained}), above every off-the-shelf retriever.
Training on the same number of pairs from MS MARCO, ReasonIR data, or SPECTER data gives at most 0.13.
Qwen3-Embedding-4B also improves, from 0.17 to 0.69, and llama-embed-nemotron-8b from 0.25 to 0.70.
The best trained retriever reaches 0.70, approaching the LLM reranker at 0.81.

\begin{figure}[!b]
\centering
\includegraphics[width=\linewidth,trim=8.8 0 3 0,clip]{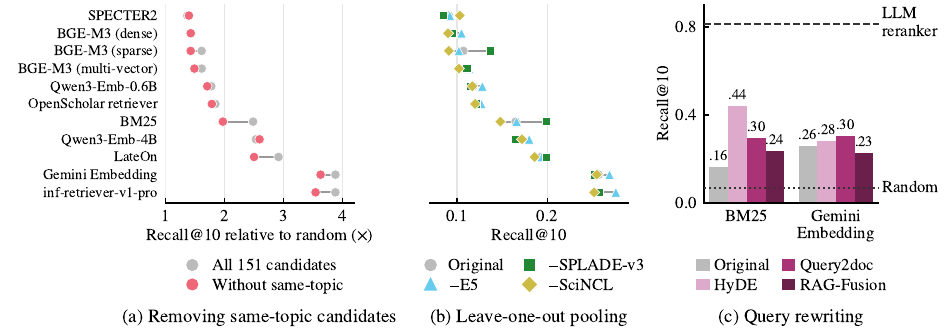}%
\caption{\textbf{Robustness of the difficulty of CPR-eval.} (a) Recall@10 divided by its random expectation in the same pool: 10/151 with all candidates, and 10/101 after removing the 50 candidates that share an OpenAlex topic with the base paper. (b) Recall@10 on the original pools and on pools rebuilt with one of the three pooling retrievers left out. (c) Recall@10 of BM25 and Gemini Embedding with the original query and with queries rewritten by GLM-5.3; the dashed line marks the LLM reranker on all 151 candidates, and the dotted line marks random ranking.}
\label{fig:robust}
\end{figure}

\subsection{Robustness Analysis}
\label{sec:difficulty}

\paragraph{Removing same-topic candidates does not make the task easier.}
A search system could simply drop papers on the query's own topic.
We simulate this by removing the 50 same-topic candidates of each query.
Because a smaller pool also raises Recall@10 by chance, we report Recall@10 relative to random ranking in each pool (10/151 and 10/101; Figure~\ref{fig:robust}a).
After the removal, retrievers reach at most 3.6 times the random rate, far below the LLM reranker at 8.5 times, computed by removing the same candidates from its ranking.

\paragraph{The difficulty does not depend on one single pooling retriever.}
Since the candidate pools come from three retrievers, E5, SPLADE-v3, and SciNCL (Section~\ref{sec:negatives}), the difficulty of CPR-eval could be an artifact of these retrievers.
Following the leave-one-out test of \citet{10.1145/290941.291014}, we rebuild the pools with each pooling retriever left out and judge the new candidates in the same way.
Across the four pools, every retriever's Recall@10 changes by at most 0.04, and the order of retrievers is largely preserved (Kendall's $\tau \geq 0.83$; Figure~\ref{fig:robust}b).

\paragraph{Query rewriting cannot fully close the gap.}
We apply HyDE~\citep{gao-etal-2023-precise}, Query2doc~\citep{wang-etal-2023-query2doc}, and RAG-Fusion~\citep{Rackauckas_2024} with GLM-5.3~\citep{glm5team2026glm5vibecodingagentic} as the generator, and retrieve with BM25 and Gemini Embedding (Figure~\ref{fig:robust}c; prompts in Appendix~\ref{sec:appendix-rewriting}).
The best combination, HyDE with BM25, raises Recall@10 from 0.16 to 0.44.
This stays below Qwen3-Embedding-4B trained on CPR-train at 0.69 and far below the LLM reranker at 0.81.

\subsection{How Far Training Generalizes}
\label{sec:generalization}

We test whether the gain from training on CPR-train carries over to unseen domains and to analogy and reasoning-intensive retrieval.

\begin{wrapfigure}{r}{0.48\linewidth}
\centering
\includegraphics[width=\linewidth]{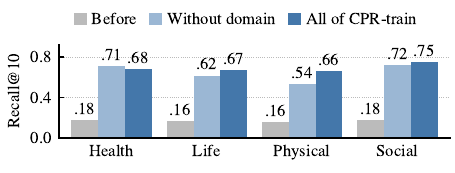}
\caption{\textbf{Recall@10 of Qwen3-Embedding-4B on domains unseen in training.} The retriever is evaluated on one OpenAlex domain before training, after training on CPR-train without that domain (5,896 pairs), and after training on all of CPR-train (10,573 pairs). Details and results for Qwen3-Embedding-0.6B are in Appendix~\ref{sec:appendix-heldout}.}
\label{fig:domain}
\vspace{-1.5\baselineskip}%
\end{wrapfigure}
\paragraph{Unseen domains.}
We hold out one of the four OpenAlex domains at a time.
Queries and positives of the held-out domain are removed from training, and the retriever is evaluated on that domain (Appendix~\ref{sec:appendix-heldout}).
Qwen3-Embedding-0.6B and 4B trained without the held-out domain still improve in every domain, raising Recall@10 by 0.21 to 0.36 and by 0.38 to 0.54 (Figure~\ref{fig:domain}).
The gain thus needs no training data from the same domain.

\paragraph{Analogy and reasoning-intensive retrieval.}
We also evaluate the trained retrievers on AnaloBench~\citep{ye-etal-2024-analobench}, which asks for the story analogous to a given one, and on the seven StackExchange subsets of BRIGHT~\citep{su2025bright}, a reasoning-intensive benchmark. Following \citet{zhao-etal-2026-rethinking}, we leave out its coding and mathematics subsets, which rely on domain-specific syntax or formal logic (Figure~\ref{fig:trained}; per-subset results in Table~\ref{tab:full-bright}).
On AnaloBench, training on CPR-train raises nDCG@10 over the untrained retriever by 0.04 to 0.12, whereas training on MS MARCO, ReasonIR data, or SPECTER data lowers it.
On BRIGHT, the gain from training on CPR-train grows with model size, from 0.05 for Qwen3-Embedding-0.6B to 0.13 for llama-embed-nemotron-8b, whereas training on the existing datasets raises nDCG@10 by at most 0.05.
In examples where only the retriever trained on CPR-train finds the relevant document, the other retrievers rank first documents that share words with the query, whereas the relevant document shares the underlying mechanism (Appendix~\ref{app:qualitative}).
Without hard negatives, most of the gain on AnaloBench disappears for the two Qwen3-Embedding retrievers, while the gain on BRIGHT remains (Appendix~\ref{sec:appendix-inbatch}).

\section{Conclusion}
\label{sec:conclusion}

We introduce cross-pollination retrieval: given a research problem, find papers on other topics whose ideas can be transferred to it. We also release CPR-Dataset, built from citations by a scalable pipeline.
On CPR-eval, existing sparse, dense, and late-interaction retrievers rarely find such papers.
Standard contrastive fine-tuning on CPR-train substantially improves retrieval.
The gain carries over to held-out domains and to analogy and reasoning-intensive retrieval, suggesting that retrievers learn relevance beyond topical similarity.
We believe that analyzing what the trained retrievers learn and using them to supply cross-field inspiration to research-assistant systems are promising directions for future work.
We hope that CPR-Dataset helps retrieval systems bring ideas from other fields to researchers and, in turn, contributes to scientific progress through cross-pollination.

\section*{Limitations}
Several limitations come from how we build CPR-Dataset.
First, positives come only from papers that the base paper cites, so a paper whose idea could be transferred but was never cited cannot become a positive.
Second, our pipeline relies on LLMs to judge analogy, transfer, and negatives; the transfer and usefulness judgments agree with a human annotator at Cohen's $\kappa$ of 0.84 to 0.92 (Appendix~\ref{app:human-agreement}), but they may still contain errors or systematic biases.
Each query is also written by an LLM from its base paper and may differ from the questions that researchers would ask.
Third, whether two papers are on different topics is decided by the OpenAlex topic classification, so a different taxonomy would yield a different set of cross-topic pairs.
Finally, we build CPR-Dataset from Semantic Scholar~\citep{kinney2023semantic} papers that meet our selection criteria, such as having full text available, and the fields, venues, and years it represents may be biased toward such papers.

Our evaluation uses a fixed pool of 151 candidates per query rather than a full corpus.
Retrieval over a full corpus is likely harder, and the Recall@10 of the LLM reranker likewise assumes a pool that already contains the positive. Our evaluation also leaves out general retrieval benchmarks, since retrieval by topical similarity is outside the scope of this work; whether training on CPR-train affects it remains open.

Our training experiments cover three retrievers, Qwen3-Embedding-0.6B and 4B and llama-embed-nemotron-8b, trained with a single contrastive objective.
Larger retrievers and other training methods may benefit from CPR-train differently.

\section*{Ethical Considerations}
Retrievers trained on CPR-train may return papers whose ideas appear transferable but do not actually help solve the given problem, so researchers should verify retrieved papers before building on them.
Because CPR-Dataset is built from papers whose full text is available, fields and venues with less available full text are underrepresented, and retrievers trained on it may serve researchers in those fields less well.
Building CPR-Dataset and training retrievers required LLM calls and GPU training, which consume energy; to avoid repeating this computation, we release CPR-Dataset and the pipeline that collects it, and will release the trained Qwen3-Embedding retrievers upon publication.
CPR-Dataset is released under ODC-BY as Semantic Scholar paper identifiers, titles, labels, and generated queries; abstracts are included only for papers whose open-access license is CC-BY, CC0, or public domain, and a script retrieves the others from Semantic Scholar.
We used AI assistants to help write and debug code for data processing, evaluation, and figures, and to edit the wording of the paper; the authors designed the research and checked all code, results, and text.

\bibliographystyle{plainnat}
\bibliography{paper}

\begin{thebibliography}{52}
\providecommand{\natexlab}[1]{#1}
\providecommand{\url}[1]{\texttt{#1}}
\expandafter\ifx\csname urlstyle\endcsname\relax
  \providecommand{\doi}[1]{doi: #1}\else
  \providecommand{\doi}{doi: \begingroup \urlstyle{rm}\Url}\fi

\bibitem[Ajith et~al.(2024)Ajith, Xia, Chevalier, Goyal, Chen, and Gao]{ajith-etal-2024-litsearch}
Anirudh Ajith, Mengzhou Xia, Alexis Chevalier, Tanya Goyal, Danqi Chen, and Tianyu Gao.
\newblock {L}it{S}earch: A retrieval benchmark for scientific literature search.
\newblock In Yaser Al-Onaizan, Mohit Bansal, and Yun-Nung Chen, editors, \emph{Proceedings of the 2024 Conference on Empirical Methods in Natural Language Processing}, pages 15068--15083, Miami, Florida, USA, November 2024. Association for Computational Linguistics.
\newblock \doi{10.18653/v1/2024.emnlp-main.840}.
\newblock URL \url{https://aclanthology.org/2024.emnlp-main.840/}.

\bibitem[Asai et~al.(2026)Asai, He, Shao, Shi, Singh, Chang, Lo, Soldaini, Feldman, D'Arcy, Wadden, Latzke, Sparks, Hwang, Kishore, Tian, Ji, Liu, Tong, Wu, Xiong, Zettlemoyer, Neubig, Weld, Downey, Yih, Koh, and Hajishirzi]{asai2024openscholarsynthesizingscientificliterature}
Akari Asai, Jacqueline He, Rulin Shao, Weijia Shi, Amanpreet Singh, Joseph~Chee Chang, Kyle Lo, Luca Soldaini, Sergey Feldman, Mike D'Arcy, David Wadden, Matt Latzke, Jenna Sparks, Jena~D. Hwang, Varsha Kishore, Minyang Tian, Pan Ji, Shengyan Liu, Hao Tong, Bohao Wu, Yanyu Xiong, Luke Zettlemoyer, Graham Neubig, Daniel~S. Weld, Doug Downey, Wen-tau Yih, Pang~Wei Koh, and Hannaneh Hajishirzi.
\newblock Synthesizing scientific literature with retrieval-augmented language models.
\newblock \emph{Nature}, 650\penalty0 (8103):\penalty0 857--863, 2026.
\newblock \doi{10.1038/s41586-025-10072-4}.
\newblock URL \url{https://doi.org/10.1038/s41586-025-10072-4}.

\bibitem[Babakhin et~al.(2025)Babakhin, Osmulski, Ak, Moreira, Xu, Schifferer, Liu, and Oldridge]{babakhin2025llamaembednemotron8buniversaltextembedding}
Yauhen Babakhin, Radek Osmulski, Ronay Ak, Gabriel Moreira, Mengyao Xu, Benedikt Schifferer, Bo~Liu, and Even Oldridge.
\newblock Llama-embed-nemotron-8b: A universal text embedding model for multilingual and cross-lingual tasks, 2025.
\newblock URL \url{https://arxiv.org/abs/2511.07025}.

\bibitem[Bajaj et~al.(2018)Bajaj, Campos, Craswell, Deng, Gao, Liu, Majumder, McNamara, Mitra, Nguyen, Rosenberg, Song, Stoica, Tiwary, and Wang]{DBLP:journals/corr/NguyenRSGTMD16}
Payal Bajaj, Daniel Campos, Nick Craswell, Li~Deng, Jianfeng Gao, Xiaodong Liu, Rangan Majumder, Andrew McNamara, Bhaskar Mitra, Tri Nguyen, Mir Rosenberg, Xia Song, Alina Stoica, Saurabh Tiwary, and Tong Wang.
\newblock {MS MARCO}: A human generated machine reading comprehension dataset, 2018.
\newblock URL \url{https://arxiv.org/abs/1611.09268}.

\bibitem[Chan et~al.(2018)Chan, Chang, Hope, Shahaf, and Kittur]{10.1145/3274300}
Joel Chan, Joseph~Chee Chang, Tom Hope, Dafna Shahaf, and Aniket Kittur.
\newblock {SOLVENT}: A mixed initiative system for finding analogies between research papers.
\newblock \emph{Proc. ACM Hum.-Comput. Interact.}, 2\penalty0 (CSCW), November 2018.
\newblock \doi{10.1145/3274300}.
\newblock URL \url{https://doi.org/10.1145/3274300}.

\bibitem[Chen et~al.(2024)Chen, Xiao, Zhang, Luo, Lian, and Liu]{chen-etal-2024-m3}
Jianlyu Chen, Shitao Xiao, Peitian Zhang, Kun Luo, Defu Lian, and Zheng Liu.
\newblock {M}3-embedding: Multi-linguality, multi-functionality, multi-granularity text embeddings through self-knowledge distillation.
\newblock In Lun-Wei Ku, Andre Martins, and Vivek Srikumar, editors, \emph{Findings of the Association for Computational Linguistics: ACL 2024}, pages 2318--2335, Bangkok, Thailand, August 2024. Association for Computational Linguistics.
\newblock \doi{10.18653/v1/2024.findings-acl.137}.
\newblock URL \url{https://aclanthology.org/2024.findings-acl.137/}.

\bibitem[Chen et~al.(2025)Chen, Liu, Zhang, Sun, Ma, Yang, Shi, Mao, and Yin]{10.1145/3696410.3714863}
Yiqun Chen, Qi~Liu, Yi~Zhang, Weiwei Sun, Xinyu Ma, Wei Yang, Daiting Shi, Jiaxin Mao, and Dawei Yin.
\newblock {TourRank}: Utilizing large language models for documents ranking with a tournament-inspired strategy.
\newblock In \emph{Proceedings of the ACM on Web Conference 2025}, WWW '25, page 1638–1652, New York, NY, USA, 2025. Association for Computing Machinery.
\newblock ISBN 9798400712746.
\newblock \doi{10.1145/3696410.3714863}.
\newblock URL \url{https://doi.org/10.1145/3696410.3714863}.

\bibitem[Cohan et~al.(2020)Cohan, Feldman, Beltagy, Downey, and Weld]{cohan-etal-2020-specter}
Arman Cohan, Sergey Feldman, Iz~Beltagy, Doug Downey, and Daniel Weld.
\newblock {SPECTER}: Document-level representation learning using citation-informed transformers.
\newblock In Dan Jurafsky, Joyce Chai, Natalie Schluter, and Joel Tetreault, editors, \emph{Proceedings of the 58th Annual Meeting of the Association for Computational Linguistics}, pages 2270--2282, Online, July 2020. Association for Computational Linguistics.
\newblock \doi{10.18653/v1/2020.acl-main.207}.
\newblock URL \url{https://aclanthology.org/2020.acl-main.207/}.

\bibitem[Cormack et~al.(2009)Cormack, Clarke, and B{\"u}ttcher]{cormack2009rrf}
Gordon~V. Cormack, Charles L.~A. Clarke, and Stefan B{\"u}ttcher.
\newblock Reciprocal rank fusion outperforms {Condorcet} and individual rank learning methods.
\newblock In \emph{Proceedings of the 32nd International ACM SIGIR Conference on Research and Development in Information Retrieval}, pages 758--759, 2009.
\newblock \doi{10.1145/1571941.1572114}.

\bibitem[Gao et~al.(2023)Gao, Ma, Lin, and Callan]{gao-etal-2023-precise}
Luyu Gao, Xueguang Ma, Jimmy Lin, and Jamie Callan.
\newblock Precise zero-shot dense retrieval without relevance labels.
\newblock In Anna Rogers, Jordan Boyd-Graber, and Naoaki Okazaki, editors, \emph{Proceedings of the 61st Annual Meeting of the Association for Computational Linguistics (Volume 1: Long Papers)}, pages 1762--1777, Toronto, Canada, July 2023. Association for Computational Linguistics.
\newblock \doi{10.18653/v1/2023.acl-long.99}.
\newblock URL \url{https://aclanthology.org/2023.acl-long.99/}.

\bibitem[Gentner(1983)]{gentner1983structure}
Dedre Gentner.
\newblock Structure-mapping: A theoretical framework for analogy.
\newblock \emph{Cognitive Science}, 7\penalty0 (2):\penalty0 155--170, 1983.
\newblock \doi{10.1207/s15516709cog0702_3}.
\newblock URL \url{https://onlinelibrary.wiley.com/doi/abs/10.1207/s15516709cog0702_3}.

\bibitem[{GLM-5-Team} et~al.(2026){GLM-5-Team}, Zeng, Lv, Hou, Du, Zheng, Chen, Yin, Ge, Huang, Xie, Zhu, Yin, Wang, Pan, Zeng, Zhang, Wang, Chen, Zhang, Jiao, Guo, Wang, Du, Wu, Wang, Li, Fan, Zhong, Liu, Zhao, Du, Dong, Lu, Shuang-Li, Cao, Liu, Jiang, Chen, Zhang, Huang, Dong, Xu, Wei, An, Niu, Zhu, Wen, Cen, Bai, Qiao, Wang, Wang, Zhu, Liu, Li, Wang, Wen, Huang, Cai, Yu, Li, Hu, Zhang, Zhang, Lin, Yang, Wang, Ai, Zhu, Yi, Chen, Wen, Sun, Zhao, Hu, Zhang, Liu, Zhang, Peng, Tai, Zhang, Liu, Wang, Yan, Ge, Liu, Chu, Zhao, Wang, Zhao, Ren, Wang, Zhang, Gui, Zhao, Li, An, Li, Yuan, Du, Liu, Zhi, Duan, Zhou, Wei, Wang, Luo, Zhang, Sha, Xu, Wu, Ding, Chen, Li, Lin, Ta, Zou, Song, Yang, Tu, Yang, Wu, Zhang, Li, Li, Fan, Qin, Tian, Zhang, Yu, Liang, Kuang, Cheng, Li, Yan, Hu, Ling, Fan, Xia, Zhang, Zhang, Pan, Zou, Zhang, Liu, Wu, Li, Wang, Zhu, Tan, Zhou, Pan, Zhang, Su, Geng, Yan, Tan, Bi, Shen, Yang, Li, Liu, Wang, Li, Wu, Zhang, Duan, Zhang, Liu, Jiang, Yan, Zhang, Wei, Chen, Feng, Yao, Chai, Wang, Zhang, Xu, Huang, Wang, Li, Dong, and Tang]{glm5team2026glm5vibecodingagentic}
{GLM-5-Team}, Aohan Zeng, Xin Lv, Zhenyu Hou, Zhengxiao Du, Qinkai Zheng, Bin Chen, Da~Yin, Chendi Ge, Chenghua Huang, Chengxing Xie, Chenzheng Zhu, Congfeng Yin, Cunxiang Wang, Gengzheng Pan, Hao Zeng, Haoke Zhang, Haoran Wang, Huilong Chen, Jiajie Zhang, Jian Jiao, Jiaqi Guo, Jingsen Wang, Jingzhao Du, Jinzhu Wu, Kedong Wang, Lei Li, Lin Fan, Lucen Zhong, Mingdao Liu, Mingming Zhao, Pengfan Du, Qian Dong, Rui Lu, Shuang-Li, Shulin Cao, Song Liu, Ting Jiang, Xiaodong Chen, Xiaohan Zhang, Xuancheng Huang, Xuezhen Dong, Yabo Xu, Yao Wei, Yifan An, Yilin Niu, Yitong Zhu, Yuanhao Wen, Yukuo Cen, Yushi Bai, Zhongpei Qiao, Zihan Wang, Zikang Wang, Zilin Zhu, Ziqiang Liu, Zixuan Li, Bojie Wang, Bosi Wen, Can Huang, Changpeng Cai, Chao Yu, Chen Li, Chengwei Hu, Chenhui Zhang, Dan Zhang, Daoyan Lin, Dayong Yang, Di~Wang, Ding Ai, Erle Zhu, Fangzhou Yi, Feiyu Chen, Guohong Wen, Hailong Sun, Haisha Zhao, Haiyi Hu, Hanchen Zhang, Hanrui Liu, Hanyu Zhang, Hao Peng, Hao Tai, Haobo Zhang, He~Liu, Hongwei Wang, Hongxi Yan, Hongyu Ge, Huan Liu, Huanpeng Chu, Jia'ni Zhao, Jiachen Wang, Jiajing Zhao, Jiamin Ren, Jiapeng Wang, Jiaxin Zhang, Jiayi Gui, Jiayue Zhao, Jijie Li, Jing An, Jing Li, Jingwei Yuan, Jinhua Du, Jinxin Liu, Junkai Zhi, Junwen Duan, Kaiyue Zhou, Kangjian Wei, Ke~Wang, Keyun Luo, Laiqiang Zhang, Leigang Sha, Liang Xu, Lindong Wu, Lintao Ding, Lu~Chen, Minghao Li, Nianyi Lin, Pan Ta, Qiang Zou, Rongjun Song, Ruiqi Yang, Shangqing Tu, Shangtong Yang, Shaoxiang Wu, Shengyan Zhang, Shijie Li, Shuang Li, Shuyi Fan, Wei Qin, Wei Tian, Weining Zhang, Wenbo Yu, Wenjie Liang, Xiang Kuang, Xiangmeng Cheng, Xiangyang Li, Xiaoquan Yan, Xiaowei Hu, Xiaoying Ling, Xing Fan, Xingye Xia, Xinyuan Zhang, Xinze Zhang, Xirui Pan, Xu~Zou, Xunkai Zhang, Yadi Liu, Yandong Wu, Yanfu Li, Yidong Wang, Yifan Zhu, Yijun Tan, Yilin Zhou, Yiming Pan, Ying Zhang, Yinpei Su, Yipeng Geng, Yong Yan, Yonglin Tan, Yuean Bi, Yuhan Shen, Yuhao Yang, Yujiang Li, Yunan Liu, Yunqing Wang, Yuntao Li, Yurong Wu, Yutao Zhang, Yuxi Duan, Yuxuan Zhang, Zezhen Liu, Zhengtao Jiang, Zhenhe Yan, Zheyu Zhang, Zhixiang Wei, Zhuo Chen, Zhuoer Feng, Zijun Yao, Ziwei Chai, Ziyuan Wang, Zuzhou Zhang, Bin Xu, Minlie Huang, Hongning Wang, Juanzi Li, Yuxiao Dong, and Jie Tang.
\newblock {GLM-5}: From vibe coding to agentic engineering, 2026.
\newblock URL \url{https://arxiv.org/abs/2602.15763}.

\bibitem[Goodfellow et~al.(2014)Goodfellow, Pouget-Abadie, Mirza, Xu, Warde-Farley, Ozair, Courville, and Bengio]{goodfellow2014gan}
Ian~J. Goodfellow, Jean Pouget-Abadie, Mehdi Mirza, Bing Xu, David Warde-Farley, Sherjil Ozair, Aaron Courville, and Yoshua Bengio.
\newblock Generative adversarial nets.
\newblock In \emph{Advances in Neural Information Processing Systems}, volume~27. Curran Associates, Inc., 2014.
\newblock URL \url{https://proceedings.neurips.cc/paper_files/paper/2014/file/f033ed80deb0234979a61f95710dbe25-Paper.pdf}.

\bibitem[Izacard et~al.(2022)Izacard, Caron, Hosseini, Riedel, Bojanowski, Joulin, and Grave]{izacard2022unsupervised}
Gautier Izacard, Mathilde Caron, Lucas Hosseini, Sebastian Riedel, Piotr Bojanowski, Armand Joulin, and Edouard Grave.
\newblock Unsupervised dense information retrieval with contrastive learning.
\newblock \emph{Transactions on Machine Learning Research}, 2022.
\newblock ISSN 2835-8856.
\newblock URL \url{https://openreview.net/forum?id=jKN1pXi7b0}.

\bibitem[Jarzynski(1997)]{jarzynski1997equilibrium}
C.~Jarzynski.
\newblock Equilibrium free-energy differences from nonequilibrium measurements: A master-equation approach.
\newblock \emph{Phys. Rev. E}, 56:\penalty0 5018--5035, Nov 1997.
\newblock \doi{10.1103/PhysRevE.56.5018}.
\newblock URL \url{https://link.aps.org/doi/10.1103/PhysRevE.56.5018}.

\bibitem[Jiayang et~al.(2023)Jiayang, Qiu, Chan, Fang, Wang, Chan, Ru, Guo, Zhang, Song, Zhang, and Zhang]{jiayang-etal-2023-storyanalogy}
Cheng Jiayang, Lin Qiu, Tsz Chan, Tianqing Fang, Weiqi Wang, Chunkit Chan, Dongyu Ru, Qipeng Guo, Hongming Zhang, Yangqiu Song, Yue Zhang, and Zheng Zhang.
\newblock {S}tory{A}nalogy: Deriving story-level analogies from large language models to unlock analogical understanding.
\newblock In Houda Bouamor, Juan Pino, and Kalika Bali, editors, \emph{Proceedings of the 2023 Conference on Empirical Methods in Natural Language Processing}, pages 11518--11537, Singapore, December 2023. Association for Computational Linguistics.
\newblock \doi{10.18653/v1/2023.emnlp-main.706}.
\newblock URL \url{https://aclanthology.org/2023.emnlp-main.706/}.

\bibitem[Jones(2009)]{jones2009burden}
Benjamin~F. Jones.
\newblock The burden of knowledge and the ``death of the renaissance man'': Is innovation getting harder?
\newblock \emph{The Review of Economic Studies}, 76\penalty0 (1):\penalty0 283--317, 2009.
\newblock \doi{10.1111/j.1467-937X.2008.00531.x}.

\bibitem[Karpukhin et~al.(2020)Karpukhin, Oguz, Min, Lewis, Wu, Edunov, Chen, and Yih]{karpukhin-etal-2020-dense}
Vladimir Karpukhin, Barlas Oguz, Sewon Min, Patrick Lewis, Ledell Wu, Sergey Edunov, Danqi Chen, and Wen-tau Yih.
\newblock Dense passage retrieval for open-domain question answering.
\newblock In Bonnie Webber, Trevor Cohn, Yulan He, and Yang Liu, editors, \emph{Proceedings of the 2020 Conference on Empirical Methods in Natural Language Processing (EMNLP)}, pages 6769--6781, Online, November 2020. Association for Computational Linguistics.
\newblock \doi{10.18653/v1/2020.emnlp-main.550}.
\newblock URL \url{https://aclanthology.org/2020.emnlp-main.550/}.

\bibitem[Kim et~al.(2026)Kim, Lee, Liu, Ko, Shao, Kim, Neubig, Koh, Chowdhery, Asai, Khattab, Choi, Kim, and Finn]{kim2026scholarcatalystbenchmarkretrievingpapers}
Sohyeon Kim, Yoonho Lee, Bo~Liu, Dayoon Ko, Rulin Shao, Seungone Kim, Graham Neubig, Pang~Wei Koh, Aakanksha Chowdhery, Akari Asai, Omar Khattab, Yejin Choi, Gunhee Kim, and Chelsea Finn.
\newblock {ScholarCatalyst}: A benchmark for retrieving papers that inspire new research, 2026.
\newblock URL \url{https://arxiv.org/abs/2610.02202}.

\bibitem[Kinney et~al.(2025)Kinney, Anastasiades, Authur, Beltagy, Bragg, Buraczynski, Cachola, Candra, Chandrasekhar, Cohan, Crawford, Downey, Dunkelberger, Etzioni, Evans, Feldman, Gorney, Graham, Hu, Huff, King, Kohlmeier, Kuehl, Langan, Lin, Liu, Lo, Lochner, MacMillan, Murray, Newell, Rao, Rohatgi, Sayre, Shen, Singh, Soldaini, Subramanian, Tanaka, Wade, Wagner, Wang, Wilhelm, Wu, Yang, Zamarron, Zuylen, and Weld]{kinney2023semantic}
Rodney Kinney, Chloe Anastasiades, Russell Authur, Iz~Beltagy, Jonathan Bragg, Alexandra Buraczynski, Isabel Cachola, Stefan Candra, Yoganand Chandrasekhar, Arman Cohan, Miles Crawford, Doug Downey, Jason Dunkelberger, Oren Etzioni, Rob Evans, Sergey Feldman, Joseph Gorney, David Graham, Fangzhou Hu, Regan Huff, Daniel King, Sebastian Kohlmeier, Bailey Kuehl, Michael Langan, Daniel Lin, Haokun Liu, Kyle Lo, Jaron Lochner, Kelsey MacMillan, Tyler Murray, Chris Newell, Smita Rao, Shaurya Rohatgi, Paul Sayre, Zejiang Shen, Amanpreet Singh, Luca Soldaini, Shivashankar Subramanian, Amber Tanaka, Alex~D. Wade, Linda Wagner, Lucy~Lu Wang, Chris Wilhelm, Caroline Wu, Jiangjiang Yang, Angele Zamarron, Madeleine~Van Zuylen, and Daniel~S. Weld.
\newblock The semantic scholar open data platform, 2025.
\newblock URL \url{https://arxiv.org/abs/2301.10140}.

\bibitem[Kulikowski(2026)]{kulikowski2026problemdifferentfieldcrossdomain}
Eryk Kulikowski.
\newblock Same problem, different field: Cross-domain solution import via domain-stripped computational fingerprints.
\newblock In \emph{Proceedings of the 2026 ACM/IEEE Joint Conference on Digital Libraries (JCDL)}, Frisco, TX, USA, 2026.
\newblock URL \url{https://arxiv.org/abs/2609.07595}.

\bibitem[Larivi{\`e}re et~al.(2015)Larivi{\`e}re, Haustein, and B{\"o}rner]{lariviere2015longdistance}
Vincent Larivi{\`e}re, Stefanie Haustein, and Katy B{\"o}rner.
\newblock Long-distance interdisciplinarity leads to higher scientific impact.
\newblock \emph{PLoS ONE}, 10\penalty0 (3):\penalty0 e0122565, 2015.
\newblock \doi{10.1371/journal.pone.0122565}.
\newblock URL \url{https://doi.org/10.1371/journal.pone.0122565}.

\bibitem[Lassance et~al.(2024)Lassance, Déjean, Formal, and Clinchant]{lassance2024spladev3newbaselinessplade}
Carlos Lassance, Hervé Déjean, Thibault Formal, and Stéphane Clinchant.
\newblock {SPLADE-v3}: New baselines for {SPLADE}, 2024.
\newblock URL \url{https://arxiv.org/abs/2403.06789}.

\bibitem[Lee et~al.(2025)Lee, Chen, Dua, Cer, Shanbhogue, Naim, Ábrego, Li, Chen, Vera, Ren, Zhang, Salz, Boratko, Han, Chen, Huang, Rao, Suganthan, Han, Doumanoglou, Gupta, Moiseev, Yip, Jain, Baumgartner, Shahi, Gomez, Mariserla, Choi, Shah, Goenka, Chen, Xia, Chen, Duddu, Chen, Walker, Zhou, Ghiya, Gleicher, Gill, Dong, Seyedhosseini, Sung, Hoffmann, and Duerig]{lee2025geminiembeddinggeneralizableembeddings}
Jinhyuk Lee, Feiyang Chen, Sahil Dua, Daniel Cer, Madhuri Shanbhogue, Iftekhar Naim, Gustavo~Hernández Ábrego, Zhe Li, Kaifeng Chen, Henrique~Schechter Vera, Xiaoqi Ren, Shanfeng Zhang, Daniel Salz, Michael Boratko, Jay Han, Blair Chen, Shuo Huang, Vikram Rao, Paul Suganthan, Feng Han, Andreas Doumanoglou, Nithi Gupta, Fedor Moiseev, Cathy Yip, Aashi Jain, Simon Baumgartner, Shahrokh Shahi, Frank~Palma Gomez, Sandeep Mariserla, Min Choi, Parashar Shah, Sonam Goenka, Ke~Chen, Ye~Xia, Koert Chen, Sai Meher~Karthik Duddu, Yichang Chen, Trevor Walker, Wenlei Zhou, Rakesh Ghiya, Zach Gleicher, Karan Gill, Zhe Dong, Mojtaba Seyedhosseini, Yunhsuan Sung, Raphael Hoffmann, and Tom Duerig.
\newblock Gemini embedding: Generalizable embeddings from gemini, 2025.
\newblock URL \url{https://arxiv.org/abs/2503.07891}.

\bibitem[Liu et~al.(2026)Liu, Yang, Xie, Ni, Gao, Li, Tang, Ouyang, Cambria, and Zhou]{liu-etal-2026-researchbench}
Yujie Liu, Zonglin Yang, Tong Xie, Jinjie Ni, Ben Gao, Yuqiang Li, Shixiang Tang, Wanli Ouyang, Erik Cambria, and Dongzhan Zhou.
\newblock {R}esearch{B}ench: Benchmarking {LLM}s in scientific discovery via inspiration-based task decomposition.
\newblock In Maria Liakata, Viviane~P. Moreira, Jiajun Zhang, and David Jurgens, editors, \emph{Findings of the {A}ssociation for {C}omputational {L}inguistics: {ACL} 2026}, pages 13187--13207, San Diego, California, United States, July 2026. Association for Computational Linguistics.
\newblock ISBN 979-8-89176-395-1.
\newblock \doi{10.18653/v1/2026.findings-acl.644}.
\newblock URL \url{https://aclanthology.org/2026.findings-acl.644/}.

\bibitem[L{\`u}(2024)]{lu2024bm25s}
Xing~Han L{\`u}.
\newblock {BM25S}: Orders of magnitude faster lexical search via eager sparse scoring, 2024.
\newblock URL \url{https://arxiv.org/abs/2407.03618}.

\bibitem[{OpenAlex}(2026)]{openalex2024topics}
{OpenAlex}.
\newblock {OpenAlex} topics.
\newblock \url{https://help.openalex.org/hc/en-us/articles/24736129405719-Topics}, 2026.
\newblock Accessed: 2026-09-23.

\bibitem[Ostendorff et~al.(2022)Ostendorff, Rethmeier, Augenstein, Gipp, and Rehm]{ostendorff-etal-2022-neighborhood}
Malte Ostendorff, Nils Rethmeier, Isabelle Augenstein, Bela Gipp, and Georg Rehm.
\newblock Neighborhood contrastive learning for scientific document representations with citation embeddings.
\newblock In Yoav Goldberg, Zornitsa Kozareva, and Yue Zhang, editors, \emph{Proceedings of the 2022 Conference on Empirical Methods in Natural Language Processing}, pages 11670--11688, Abu Dhabi, United Arab Emirates, December 2022. Association for Computational Linguistics.
\newblock \doi{10.18653/v1/2022.emnlp-main.802}.
\newblock URL \url{https://aclanthology.org/2022.emnlp-main.802/}.

\bibitem[Pandey et~al.(2026)Pandey, Kwon, and Yu]{pandey2026musesbenchmarkprospectiveintellectualroots}
Rohan Pandey, Sunjae Kwon, and Hong Yu.
\newblock {MUSES}: A benchmark for prospective intellectual-roots retrieval, 2026.
\newblock URL \url{https://arxiv.org/abs/2609.00313}.

\bibitem[Rackauckas(2024)]{Rackauckas_2024}
Zackary Rackauckas.
\newblock {RAG-Fusion}: A new take on retrieval augmented generation.
\newblock \emph{International Journal on Natural Language Computing}, 13\penalty0 (1):\penalty0 37–47, February 2024.
\newblock ISSN 2319-4111.
\newblock \doi{10.5121/ijnlc.2024.13103}.
\newblock URL \url{http://dx.doi.org/10.5121/ijnlc.2024.13103}.

\bibitem[Robertson and Zaragoza(2009)]{10.1561/1500000019}
Stephen Robertson and Hugo Zaragoza.
\newblock The probabilistic relevance framework: {BM25} and beyond.
\newblock \emph{Found. Trends Inf. Retr.}, 3\penalty0 (4):\penalty0 333–389, April 2009.
\newblock ISSN 1554-0669.
\newblock \doi{10.1561/1500000019}.
\newblock URL \url{https://doi.org/10.1561/1500000019}.

\bibitem[Shao et~al.(2025)Shao, Qiao, Kishore, Muennighoff, Lin, Rus, Low, Min, Yih, Koh, and Zettlemoyer]{shao2025reasonir}
Rulin Shao, Rui Qiao, Varsha Kishore, Niklas Muennighoff, Xi~Victoria Lin, Daniela Rus, Bryan Kian~Hsiang Low, Sewon Min, {Wen-tau} Yih, Pang~Wei Koh, and Luke Zettlemoyer.
\newblock Reason{IR}: Training retrievers for reasoning tasks.
\newblock In \emph{Second Conference on Language Modeling}, 2025.
\newblock URL \url{https://openreview.net/forum?id=kkBCNLMbGj}.

\bibitem[Shen et~al.(2026)Shen, Druckmann, and Zou]{shen2026unlockingllmcreativityscience}
Andrew Shen, Shaul Druckmann, and James Zou.
\newblock Unlocking {LLM} creativity in science through analogical reasoning, 2026.
\newblock URL \url{https://arxiv.org/abs/2605.11258}.

\bibitem[Singh et~al.(2023)Singh, D{'}Arcy, Cohan, Downey, and Feldman]{singh-etal-2023-scirepeval}
Amanpreet Singh, Mike D{'}Arcy, Arman Cohan, Doug Downey, and Sergey Feldman.
\newblock {S}ci{R}ep{E}val: A multi-format benchmark for scientific document representations.
\newblock In Houda Bouamor, Juan Pino, and Kalika Bali, editors, \emph{Proceedings of the 2023 Conference on Empirical Methods in Natural Language Processing}, pages 5548--5566, Singapore, December 2023. Association for Computational Linguistics.
\newblock \doi{10.18653/v1/2023.emnlp-main.338}.
\newblock URL \url{https://aclanthology.org/2023.emnlp-main.338/}.

\bibitem[Sohl-Dickstein et~al.(2015)Sohl-Dickstein, Weiss, Maheswaranathan, and Ganguli]{sohldickstein2015deep}
Jascha Sohl-Dickstein, Eric Weiss, Niru Maheswaranathan, and Surya Ganguli.
\newblock Deep unsupervised learning using nonequilibrium thermodynamics.
\newblock In Francis Bach and David Blei, editors, \emph{Proceedings of the 32nd International Conference on Machine Learning}, volume~37 of \emph{Proceedings of Machine Learning Research}, pages 2256--2265, Lille, France, 07--09 Jul 2015. PMLR.
\newblock URL \url{https://proceedings.mlr.press/v37/sohl-dickstein15.html}.

\bibitem[Sourty et~al.(2026)Sourty, Chaffin, Moura~Junior, and Chatelain]{sourty2026denseonlateon}
Rapha{\"e}l Sourty, Antoine Chaffin, Paulo~Roberto Moura~Junior, and Am{\'e}lie Chatelain.
\newblock {DenseOn} with the {LateOn}: Fully open dense and late-interaction models for multilingual, long-context, and code search, 2026.
\newblock URL \url{https://arxiv.org/abs/2607.27178}.

\bibitem[Su et~al.(2025)Su, Yen, Xia, Shi, Muennighoff, Wang, Haisu, Shi, Siegel, Tang, Sun, Yoon, Arik, Chen, and Yu]{su2025bright}
Hongjin Su, Howard Yen, Mengzhou Xia, Weijia Shi, Niklas Muennighoff, {Han-yu} Wang, Liu Haisu, Quan Shi, Zachary~S Siegel, Michael Tang, Ruoxi Sun, Jinsung Yoon, Sercan~O Arik, Danqi Chen, and Tao Yu.
\newblock {BRIGHT}: A realistic and challenging benchmark for reasoning-intensive retrieval.
\newblock In \emph{The Thirteenth International Conference on Learning Representations}, 2025.
\newblock URL \url{https://openreview.net/forum?id=ykuc5q381b}.

\bibitem[Sun et~al.(2026)Sun, Long, Yang, Wang, Luo, Shen, Wang, Zhou, Guo, Wei, Wang, and Gu]{DIVER}
Duolin Sun, Meixiu Long, Dan Yang, Junjie Wang, Yecheng Luo, Yue Shen, Jian Wang, Hualei Zhou, Chunxiao Guo, Peng Wei, Jiahai Wang, and Jinjie Gu.
\newblock {DIVER}: A multi-stage approach for reasoning-intensive information retrieval, 2026.
\newblock URL \url{https://arxiv.org/abs/2508.07995}.

\bibitem[Tchuindjo et~al.(2026)Tchuindjo, Shah, and Khattab]{OBLIQ-Bench}
Diane Tchuindjo, Devavrat Shah, and Omar Khattab.
\newblock {OBLIQ-Bench}: Exposing overlooked bottlenecks in modern retrievers with latent and implicit queries, 2026.
\newblock URL \url{https://arxiv.org/abs/2605.06235}.

\bibitem[Upadhyay et~al.(2024)Upadhyay, Pradeep, Thakur, Craswell, and Lin]{upadhyay2024umbrela}
Shivani Upadhyay, Ronak Pradeep, Nandan Thakur, Nick Craswell, and Jimmy Lin.
\newblock {UMBRELA}: {UMbrela} is the (open-source reproduction of the) {Bing} {RELevance} assessor, 2024.
\newblock URL \url{https://arxiv.org/abs/2406.06519}.

\bibitem[Uzzi et~al.(2013)Uzzi, Mukherjee, Stringer, and Jones]{uzzi2013atypical}
Brian Uzzi, Satyam Mukherjee, Michael Stringer, and Ben Jones.
\newblock Atypical combinations and scientific impact.
\newblock \emph{Science}, 342\penalty0 (6157):\penalty0 468--472, 2013.
\newblock \doi{10.1126/science.1240474}.
\newblock URL \url{https://www.science.org/doi/abs/10.1126/science.1240474}.

\bibitem[van~den Oord et~al.(2019)van~den Oord, Li, and Vinyals]{oord2019representationlearningcontrastivepredictive}
Aaron van~den Oord, Yazhe Li, and Oriol Vinyals.
\newblock Representation learning with contrastive predictive coding, 2019.
\newblock URL \url{https://arxiv.org/abs/1807.03748}.

\bibitem[Voorhees and Harman(2005)]{voorhees2005trec}
Ellen~M. Voorhees and Donna~K. Harman, editors.
\newblock \emph{{TREC}: Experiment and Evaluation in Information Retrieval}.
\newblock Digital Libraries and Electronic Publishing. MIT Press, Cambridge, MA, 2005.
\newblock ISBN 9780262220736.
\newblock URL \url{https://mitpress.mit.edu/9780262220736/trec/}.

\bibitem[Wang et~al.(2023)Wang, Yang, and Wei]{wang-etal-2023-query2doc}
Liang Wang, Nan Yang, and Furu Wei.
\newblock Query2doc: Query expansion with large language models.
\newblock In Houda Bouamor, Juan Pino, and Kalika Bali, editors, \emph{Proceedings of the 2023 Conference on Empirical Methods in Natural Language Processing}, pages 9414--9423, Singapore, December 2023. Association for Computational Linguistics.
\newblock \doi{10.18653/v1/2023.emnlp-main.585}.
\newblock URL \url{https://aclanthology.org/2023.emnlp-main.585/}.

\bibitem[Wang et~al.(2024)Wang, Yang, Huang, Jiao, Yang, Jiang, Majumder, and Wei]{wang2024textembeddingsweaklysupervisedcontrastive}
Liang Wang, Nan Yang, Xiaolong Huang, Binxing Jiao, Linjun Yang, Daxin Jiang, Rangan Majumder, and Furu Wei.
\newblock Text embeddings by weakly-supervised contrastive pre-training, 2024.
\newblock URL \url{https://arxiv.org/abs/2212.03533}.

\bibitem[{Xiaomi MiMo Team}(2026)]{mimo2026v26pro}
{Xiaomi MiMo Team}.
\newblock Mimo-v2.6-pro-rl.
\newblock \url{https://huggingface.co/XiaomiMiMo/MiMo-V2.6-Pro-RL}, 2026.

\bibitem[Yang et~al.(2025)Yang, Liu, Gao, Xie, Li, Ouyang, Poria, Cambria, and Zhou]{yang2025moosechem}
Zonglin Yang, Wanhao Liu, Ben Gao, Tong Xie, Yuqiang Li, Wanli Ouyang, Soujanya Poria, Erik Cambria, and Dongzhan Zhou.
\newblock {MOOSE}-chem: Large language models for rediscovering unseen chemistry scientific hypotheses.
\newblock In \emph{The Thirteenth International Conference on Learning Representations}, 2025.
\newblock URL \url{https://openreview.net/forum?id=X9OfMNNepI}.

\bibitem[Yao et~al.(2025)Yao, Wan, Hong, Zhang, Yang, Jiang, Xu, Lu, Xu, Chu, Wang, and Qi]{infx2025retriever}
Yichen Yao, Jiahe Wan, Yuxin Hong, Mengna Zhang, Junhan Yang, Zhouyu Jiang, Qing Xu, Kuan Lu, Yinghui Xu, Wei Chu, Emma Wang, and Yuan Qi.
\newblock {INF-X-Retriever}.
\newblock \url{https://yaoyichen.github.io/INF-X-Retriever}, 2025.

\bibitem[Ye et~al.(2024)Ye, Wang, Choi, Lu, Sharma, Shen, Tiyyala, Andrews, and Khashabi]{ye-etal-2024-analobench}
Xiao Ye, Andrew Wang, Jacob Choi, Yining Lu, Shreya Sharma, Lingfeng Shen, Vijay~Murari Tiyyala, Nicholas Andrews, and Daniel Khashabi.
\newblock {A}nalo{B}ench: Benchmarking the identification of abstract and long-context analogies.
\newblock In Yaser Al-Onaizan, Mohit Bansal, and Yun-Nung Chen, editors, \emph{Proceedings of the 2024 Conference on Empirical Methods in Natural Language Processing}, pages 13060--13082, Miami, Florida, USA, November 2024. Association for Computational Linguistics.
\newblock \doi{10.18653/v1/2024.emnlp-main.725}.
\newblock URL \url{https://aclanthology.org/2024.emnlp-main.725/}.

\bibitem[Zhang et~al.(2025)Zhang, Li, Long, Zhang, Lin, Yang, Xie, Yang, Liu, Lin, Huang, and Zhou]{qwen3embedding}
Yanzhao Zhang, Mingxin Li, Dingkun Long, Xin Zhang, Huan Lin, Baosong Yang, Pengjun Xie, An~Yang, Dayiheng Liu, Junyang Lin, Fei Huang, and Jingren Zhou.
\newblock Qwen3 embedding: Advancing text embedding and reranking through foundation models, 2025.
\newblock URL \url{https://arxiv.org/abs/2506.05176}.

\bibitem[Zhao et~al.(2026)Zhao, Wei, Song, Zhang, Zhao, and Cohan]{zhao-etal-2026-rethinking}
Yilun Zhao, Jinbiao Wei, Tingyu Song, Siyue Zhang, Chen Zhao, and Arman Cohan.
\newblock Rethinking reasoning-intensive retrieval: Evaluating and advancing retrievers in agentic search systems.
\newblock In Maria Liakata, Viviane~P. Moreira, Jiajun Zhang, and David Jurgens, editors, \emph{Proceedings of the 64th Annual Meeting of the {A}ssociation for {C}omputational {L}inguistics (Volume 1: Long Papers)}, pages 36776--36806, San Diego, California, United States, July 2026. Association for Computational Linguistics.
\newblock ISBN 979-8-89176-390-6.
\newblock \doi{10.18653/v1/2026.acl-long.1705}.
\newblock URL \url{https://aclanthology.org/2026.acl-long.1705/}.

\bibitem[Zobel(1998)]{10.1145/290941.291014}
Justin Zobel.
\newblock How reliable are the results of large-scale information retrieval experiments?
\newblock In \emph{Proceedings of the 21st Annual International ACM SIGIR Conference on Research and Development in Information Retrieval}, SIGIR '98, page 307–314, New York, NY, USA, 1998. Association for Computing Machinery.
\newblock ISBN 1581130155.
\newblock \doi{10.1145/290941.291014}.
\newblock URL \url{https://doi.org/10.1145/290941.291014}.

\end{thebibliography}

\clearpage
\appendix{

\section{Examples of Analogous Pairs}
\label{app:analogy-example}
Figure~\ref{fig:analogy-example} shows an analogous pair mined by our pipeline, represented following \citet{shen2026unlockingllmcreativityscience}: the objects of each problem, the mappings between objects that play the same role, and the relations that hold among them in both problems.
The base paper $q$ completes a partially filled Latin square, and the cited paper $d$ extends a precoloring of a graph.
Four objects of $q$ map to four objects of $d$, and three relations hold among the mapped objects in both problems.

\begin{figure}[h]
\centering
\includegraphics{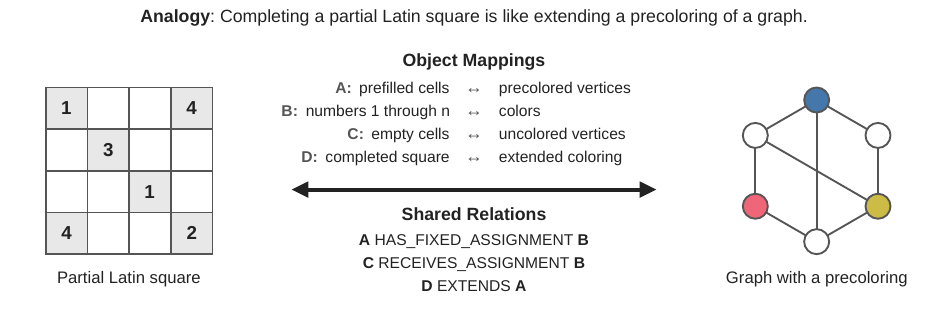}
\caption{\textbf{An analogous pair mined by our pipeline.} $q$: Solving the Latin Square Completion Problem by Memetic Graph Coloring. $d$: Precoloring Extension. I. Interval Graphs. Object mappings and shared relations are those extracted by the analogy judge (Appendix~\ref{app:prompt-analogy}), with object names shortened.}
\label{fig:analogy-example}
\end{figure}

Table~\ref{tab:more-examples} shows four more analogous pairs mined by our pipeline, each with one hard negative and two of the object mappings extracted by the analogy judge.

\begin{table}[t]
\centering
\small
\renewcommand{\arraystretch}{1.15}
\begin{tabularx}{\linewidth}{@{}>{\raggedright\arraybackslash}X>{\raggedright\arraybackslash}X>{\raggedright\arraybackslash}X>{\raggedright\arraybackslash}X@{}}
\toprule
Query & Positive & Hard negative & Object mappings (query side $\mapsto$ positive) \\
\midrule
How to define a representative conformation tensor average for turbulent viscoelastic flows that respects the non-Euclidean geometry of positive-definite tensors & A Differential Geometric Approach to the Geometric Mean of Symmetric Positive-Definite Matrices & Numerical Simulation of the Planar Extrudate Swell of Pseudoplastic and Viscoelastic Fluids & conformation tensors $\mapsto$ symmetric positive-definite matrices; geometric mean conformation tensor $\mapsto$ Riemannian geometric mean \\
\addlinespace[0.4em]
How to understand the role of personal and social factors in the development of social entrepreneurship in Pakistan & Economic Action and Social Structure: The Problem of Embeddedness & Shaker A. Zahra: Pioneering Entrepreneurship Scholar & social entrepreneurs in Pakistan $\mapsto$ individuals undertaking economic action; social capital and networks $\mapsto$ structures of social relations \\
\addlinespace[0.4em]
How to cost-effectively obtain large-scale sequence information for upland cotton flower development when the genome sequence is unavailable & An EST Resource for Tilapia Based on 17 Normalized Libraries & Improvement of Cotton Fiber Quality by Rabbit Keratin Gene Transformation & cotton normalized cDNA library $\mapsto$ tilapia normalized cDNA libraries; cotton EST reads $\mapsto$ tilapia EST reads \\
\addlinespace[0.4em]
How to understand positive and negative responses to death portrayed in animated Disney films & The Development of Coping & Frozen in Time: How Disney Gender-Stereotypes Its Most Powerful Princess & film characters facing death $\mapsto$ children facing real-life stressors; coping portrayed in the films $\mapsto$ families of coping strategies \\
\bottomrule
\end{tabularx}
\caption{\textbf{More analogous pairs mined by our pipeline.} Each hard negative shares an OpenAlex topic with the base paper and was rated 1 (related but not usable) by the judge of Step N2. Two of the object mappings extracted by the analogy judge (Appendix~\ref{app:prompt-analogy}) are shown for each pair; queries, titles, and object names are shortened.}
\label{tab:more-examples}
\end{table}

\section{Prompts}
\label{sec:appendix-prompts}
The LLM steps of the pipeline use GPT-5.6 Sol, GPT-6.1 Sol, and GPT-6 Astra, all accessed through the OpenAI API in October 2026.
\subsection{Analogy Judgment}\label{app:prompt-analogy}%
We use GPT-5.6 Sol (reasoning off) for the analogy judgment.
For each candidate pair, the input consists of the title and abstract of the problem paper $q$, the title and abstract of the cited paper $d$, the sentences in $q$ that cite $d$ (the relevance text), and the highest-ranked OpenAlex topic of each paper (the domains in the prompt).
A pair passes when the decision is \texttt{observed\_analogy} or \texttt{potential\_analogy}.
The full prompt is shown below; placeholders in braces are filled per pair.

\begin{promptbox}
You are judging whether a cited solution D forms a useful cross-domain analogy for a problem paper q.

Your task combines two formerly separate checks:
1. Extract and validate the structural analogy between q and D.
2. Decide whether q actually transfers D's contribution and, separately, whether the analogy could plausibly help q even when actual transfer is not evidenced.

Use only the supplied problem, solution, and relevance text as evidence. Domain labels are screening metadata, not sufficient evidence by themselves. Do not rely on outside knowledge. Do not invent mappings, relations, uses, or benefits.

Definitions

- q / problem domain: the paper or task seeking help. This is the source side of every object mapping.
- D / solution domain: the cited paper whose structure or contribution may be transferred. This is the target side of every object mapping.
- Functional object mapping: a correspondence between a q-side object and a D-side object based on what each does in the relevant system, not merely on a shared name, topic, or attribute.
- Shared relation: the same predicate that holds among mapped objects in both domains. Every argument in a shared relation must occur in object_mappings. A property of one object, such as "is liquid" or "is large", is not a shared relation.
- Structural analogy: at least two grounded functional object mappings and at least one grounded shared relation, with the same relational pattern holding across the two domains.
- Actual transfer: the relevance text shows that q uses, adapts, or builds on D's contribution to address q's problem through a validated cross-domain relational correspondence. Adoption alone is not evidence of analogy: it is not enough that q takes something from D; the evidence must show that D's relational structure corresponds to and is transferred into a different problem structure in q.
- Analogy-mediated potential: even if actual transfer is not shown, carrying a concrete contribution from D to q along the validated mappings would plausibly help address q's stated goal. Generic usefulness, topical relatedness, or direct reuse without a cross-domain mapping is not analogy-mediated potential.

Joint-evidence rule

- The Relevance text does not need to state every object mapping or shared relation explicitly. You may combine the Cited Solution D's Description with the Relevance text and the Problem q text: the Description may establish D's contribution, objects, and internal relations, while the Relevance text may establish how q invokes, uses, adapts, or connects to that contribution. If these supplied texts jointly ground at least two functional object mappings and one shared relation between q and D, the structural analogy may be accepted and the pair may receive a positive decision under the final decision rules.
- This rule does not permit outside knowledge or invented correspondences. The Relevance text must identify a concrete connection between q and D rather than merely cite D in a background list. For observed_analogy, the Relevance text must still evidence actual use, adaptation, or building on D's contribution through the validated correspondence. If actual transfer is not evidenced, a positive decision is limited to potential_analogy and still requires a concrete, evidence-grounded transfer_path that addresses q's stated goal.

Apply these exclusions before accepting an analogy

1. Same problem: Set same_problem to true when D is prior work on the same task as q, judged by what the two papers set out to do and by D's role in the relevance text. Different terminology or domain labels do not by themselves make the pair cross-domain. A same-problem pair is not an analogy.
2. Background or narrative only: A citation used only for motivation, a literature list, historical context, contrast, or evidence is not an analogy unless q transfers a relational structure from D.
3. Tool or data as-is: Using D only as an unchanged dataset, benchmark, software package, pretrained resource, measurement tool, or lookup source is not an analogy. Adaptation of the resource counts only when the adaptation itself is grounded through the mappings and addresses q's goal.
4. Term borrowing: Reusing a name or metaphor without a functional relational correspondence is not an analogy.
5. Surface similarity: Shared topics, nouns, attributes, or broad statements such as "both involve networks" are insufficient.
6. Unsupported possibility: The fact that D could conceivably inspire q is insufficient. Potential requires a specific, evidence-grounded transfer path from a contribution of D, through the mappings, to a difficulty or objective in q.
7. Unconnected mapping: A plausible mapping that does not address what q is trying to achieve is neither actual transfer nor useful potential.
8. Adoption without relational transfer: The fact that q adopts, imports, applies, or modifies something from D does not by itself establish an analogy. Reject the pair unless the supplied evidence grounds how D's relational structure corresponds to a different problem structure in q and is transferred through that correspondence.

Judge in this order

A. Determine whether q and D address the same problem. If they do, reject the analogy.
B. Propose only functional object mappings explicitly stated or jointly and strongly implied by the Problem q text, the Cited Solution D's Description, and the Relevance text. Prefer 4-8 mappings when the evidence supports them, but never add mappings merely to reach a count.
C. Test each shared relation. Use predicate form and the same predicate name on both sides. Confirm that every predicate argument appears in object_mappings.
D. Try to falsify the analogy using the exclusions above. In particular, distinguish relational transfer from background citation, unchanged tool/data use, term borrowing, and surface resemblance.
E. Judge actual transfer from the relevance text. transfer_holds is true only when q demonstrably uses, adapts, or builds on D's method, formulation, evaluation design, data construction, or finding to address q's problem through the validated cross-domain relational correspondence. Mere adoption from D does not establish transfer_holds. Statements about what q could do do not establish actual transfer.
F. Judge potential independently. transferable is true only when there is a concrete transfer_path explaining: (i) what contribution comes from D, (ii) how the validated mappings carry it into q, and (iii) how that transfer could help q's stated goal. transfer_holds=true requires transferable=true.
G. Assign the final decision by the deterministic rules below.

Final decision rules

- "observed_analogy": structural_analogy=true and transfer_holds=true.
- "potential_analogy": structural_analogy=true, transfer_holds=false, and transferable=true.
- "not_analogy": every other case.

If same_problem=true, or if the pair is only background, narrative, tool/data as-is, term borrowing, or surface similarity, return empty object_mappings and shared_relations, set structural_analogy, transfer_holds, and transferable to false, use an empty transfer_path, and choose not_analogy.

transfer_type must be exactly one of:
- "method"
- "problem_formulation"
- "evaluation_design"
- "data_construction"
- "finding_or_evidence"
- "narrative_only"
- "tool_as_is"
- "same_problem"
- "none"

Use narrative_only for a list/background/motivation mention with nothing structurally transferred; tool_as_is for unchanged resource use; same_problem when same_problem is true; and none when no more specific type applies. For observed_analogy or potential_analogy, choose the contribution type that would be transferred.

Return exactly one JSON object matching this schema and no other text:

{
  "target_domain": "the supplied solution domain",
  "same_problem": false,
  "object_mappings": [
    {
      "source": "q-side problem object",
      "target": "D-side solution object",
      "mapping_rationale": "brief evidence-grounded functional correspondence"
    }
  ],
  "shared_relations": [
    {
      "relation": "PREDICATE_NAME",
      "problem": "PREDICATE_NAME(q_object_A, q_object_B)",
      "solution": "PREDICATE_NAME(D_object_A, D_object_B)"
    }
  ],
  "structural_analogy": true,
  "transfer_holds": false,
  "transferable": true,
  "transfer_type": "method",
  "transfer_path": "D contribution -> mapped q structure -> expected help for q's goal",
  "critic": {
    "background_only": false,
    "tool_or_data_as_is": false,
    "term_borrowing_only": false,
    "surface_similarity_only": false,
    "mappings_grounded": true,
    "relations_grounded": true,
    "relation_arguments_mapped": true,
    "addresses_problem_goal": true
  },
  "decision": "potential_analogy",
  "reason": "one or two concise sentences citing or closely paraphrasing the decisive supplied evidence"
}

Boolean fields must be JSON booleans, not strings. Do not include markdown fences. Do not reveal hidden reasoning; provide only the requested concise rationales and final JSON fields.

Problem Domain: {source_domain}
Solution Domain: {target_domain}

Problem q:
{problem}

Cited Solution D:
- Title: {solution_title}
- Description: {description}
- Relevance: {relevance}
\end{promptbox}
\subsection{Transfer Judgment}\label{app:prompt-transfer}%
Pairs that pass the analogy judgment are checked again by a separate LLM, GPT-6.1 Sol (low reasoning effort).
The input consists of the titles and abstracts of $q$ and $d$, the sentences in $q$ that cite $d$, and the output of the analogy judgment (decision, object mappings, shared relations, transfer type, transfer path, and reason).
A pair is kept when the verdict is \texttt{transfer}.

\begin{promptbox}
You are checking a candidate analogy between a problem paper q and a cited paper D.
A first-pass judge has already proposed object mappings, shared relations and a transfer path between them.
Decide whether something carried over from D actually helps solve q's problem.

The pair passes (transfer) when all of these hold:
1. q and D address different problems. D is not prior work on the same task as q.
2. q actually uses, adapts or builds on a contribution of D (a method, formulation, component, measure,
   finding, design or resource) for its own problem, as shown by the supplied texts.
3. The proposed mappings that connect D's contribution to q's problem are supported by the supplied texts.
4. What is carried over serves what q is trying to achieve.

It does not matter whether the contribution is general-purpose or already common practice: a widely used
technique still counts if it was carried over from D's different problem and helps q.

Reject the pair if any of these holds:
- same_problem: D already addresses q's problem (same task, or the same problem line in another dataset,
  population, language or setting).
- background_only: q cites D only for motivation, related work, contrast or history, without using it.
- unsupported_mappings: the proposed mappings or transfer are not supported by the supplied texts.

Use only the supplied texts. Do not rely on outside knowledge.

Problem paper q
Title: {q_title}
Abstract: {q_abstract}

Cited paper D
Title: {d_title}
Abstract: {d_abstract}

Sentences in q that cite D:
{citing_sentences}

First-pass judgment
Decision: {decision}
Object mappings (q -> D): {mappings}
Shared relations (q <-> D): {relations}
Transfer type: {transfer_type}
Transfer path: {transfer_path}
Reason: {reason}

Return JSON only:
{"verdict": "transfer" | "reject",
 "reject_reason": "same_problem" | "background_only" | "unsupported_mappings" | null,
 "reason": "one or two sentences citing the decisive evidence"}
\end{promptbox}
\subsection{Query Generation}\label{app:prompt-query}%
Queries are generated by GPT-6 Astra in three steps.
(1)~\emph{Problem extraction.} From the title and abstract of $q$ and the sentences in $q$ that cite $d$ (with the citation replaced by \texttt{[CITATION]}), the model states the problem that $q$ is solving at the point of the citation. The prompt also gives what $q$ took from $d$ (the first part of the transfer path from the analogy judgment, Appendix~\ref{app:prompt-analogy}) and tells the model not to mention it.
(2)~\emph{Rewriting.} The problem is rewritten as a ``How to \ldots'' query that keeps only the goal and removes the terminology of $d$'s field(s), which fill the placeholder \texttt{\{base\_domain\}}.
(3)~\emph{Review.} A separate call checks the query (\texttt{problem\_only}, \texttt{grounded}, \texttt{well\_formed}), and simple rules flag citation markers, uppercase tokens, meta-language (e.g., ``analogy'', ``mapping''), method phrases (e.g., ``by using''), queries that do not start with ``How to'', and queries longer than 45 words.
A query that fails any check is regenerated with the reviewer's objection, up to two times; pairs that still fail are dropped.

\paragraph{Problem extraction.}
\begin{promptbox}
You are given the title and abstract of a research paper, and the sentences in which this paper cites a specific prior work (the citation is marked [CITATION]).
Title: {title}
Abstract: {abstract}
Citation context:
{citation_context}
Task: state the problem the authors are trying to solve AT THE POINT of the citation context, in 1-2 sentences. Describe only the problem (what is difficult or missing and why), not the paper's method, contribution, or results. If the context is a background or related-work remark, state the problem that remark is motivating.
If the citation context describes a component of the paper's own method, do not state that component: state the need it serves, that is, the difficulty that existed before the paper's approach and that the component is meant to overcome.
State the problem only: do not describe the paper's method, model, contribution, or results, and do not use "by", "through" or "using" phrases that name a technique. Everything you write must be supported by the supplied text.
What the authors took from the cited work: {contribution}
State the problem as it stood before they had this: what they needed and why it was hard. Do not name or describe this contribution, or the way the authors applied it.
Return only the problem statement:
\end{promptbox}

\paragraph{Rewriting.}
\begin{promptbox}
Rewrite this problem as a concise search query - stating ONLY the goal, with NO hints about mechanisms or methods.
Original Problem:
{problem}
Solution domain to hide: {base_domain}
Rules:
1. State only what needs to be achieved, not the method, contribution, or result
2. Remove all terminology from the solution domain ({base_domain}) and anything that hints at how the problem is solved
3. Use only what the original problem states; add nothing
4. Keep the conditions and constraints that define this particular problem (what is studied, what is limited or required), so that a solution from another field could be recognized as fitting it; drop incidental details
5. Remove all "by...", "through...", "using..." phrases
6. Format: "How to [goal]"
Return only the rewritten problem:
\end{promptbox}

\paragraph{Review.}
\begin{promptbox}
You are checking a search query that is supposed to state a research PROBLEM (what must be achieved), written from the paper below. Judge the query strictly.
Paper title: {title}
Paper abstract: {abstract}
Sentences from the paper that cite the work the query was written for (citation markers masked as [CITATION]):
{citation_context}
Query: {query}
Answer in JSON with these fields:
- "problem_only": true if the query states only what must be achieved; false if it describes a method, model, algorithm, training procedure, or the paper's contribution or result.
- "grounded": true if everything in the query is supported by the title, abstract, or the cited sentences; false if it introduces information not in them.
- "well_formed": true if it is one natural-language question in the form "How to ...", in plain lowercase English apart from proper names, with no uppercase predicate tokens or mathematical variables, no citation markers, and no meta-language about analogies, mappings, graphs, or instructions (technical uses of words such as "domain adaptation" or "multi-domain dialogue" are fine).
- "issue": one short sentence naming the problem if any field is false, else "".
Return only the JSON.
\end{promptbox}

\paragraph{Regeneration.}
The problem-extraction prompt is reused with one added line, ``Write in plain prose without mathematical variables or symbols.'', placed before the line about what the authors took from the cited work. The following text is appended to it:
\begin{promptbox}
A previous attempt was rejected. Previous problem statement: {previous_problem}
Reviewer's objection: {issue}
Write a corrected problem statement that resolves the objection.
\end{promptbox}
and the following text is appended to the rewriting prompt:
\begin{promptbox}
A previous rewrite was rejected because: {issue} Keep the conditions of the problem that this objection names, while still stating only the goal.
\end{promptbox}

\subsection{Deduplication of Queries}\label{app:prompt-dedup}
When several queries are written from the same base paper $q$, GPT-6 Astra first compares every two queries and judges whether they ask about the same problem.
Among queries judged to ask about the same problem, it then picks the cited paper that forms the strongest analogy with $q$, and only that pair is kept.
This step runs before usefulness scoring, so it does not use $u(q,d)$.

\paragraph{Same-problem check.}
\begin{promptbox}
Two search queries were written from the same research paper, each for a different place where the paper cites prior work.

Query A: {query_a}
Query B: {query_b}

Do the two queries ask about essentially the same problem, so that a paper answering one would answer the other? Differences in wording alone do not make them different problems.

Answer in JSON only: {"same_problem": true|false, "reason": "<one sentence>"}
\end{promptbox}

\paragraph{Strongest analogy.}
\begin{promptbox}
A research paper (the problem paper) cites several prior works for the same need. For each cited work you are given its title and abstract, the sentences of the problem paper that cite it, and the analogical mapping extracted between the two papers.

Problem paper title: {q_title}
Problem paper abstract: {q_abstract}

{candidates}

Which cited work forms the strongest analogy with the problem paper? A stronger analogy transfers more of the cited work's own structure (its method, formulation or finding) to a problem in a different setting, through mappings that are central to what the problem paper does, rather than a surface resemblance or a generic tool.

Answer in JSON only: {"strongest": <number>, "reason": "<one sentence>"}
\end{promptbox}

\subsection{Usefulness Scoring for Positives}\label{app:prompt-usefulness}%
We score each remaining pair $(q,d)$ with GPT-6 Astra, using a modified version of the UMBRELA relevance prompt~\citep{upadhyay2024umbrela} that asks whether the passage could be used to solve the problem in the query rather than whether it is on the same topic.
The query is the generated query of $q$ (Appendix~\ref{app:prompt-query}), and the passage is the title and abstract of $d$ (abstract truncated to 2,000 characters).
The prompt also includes the object mappings and shared relations from the analogy judgment (Appendix~\ref{app:prompt-analogy}) as optional reference.
We keep a pair only if its final score is 2 or higher.

\begin{promptbox}
Given a query and a passage, you must provide a score on an integer scale of 0 to 3 with the following meanings: 0 = represent that the passage has nothing to do with the query, 1 = represents that the passage seems related to the query but does not provide anything that could be used to solve its problem, 2 = represents that the passage has a method, formulation or finding that could be used to solve the problem in the query, but it is partial, unclear, or has to be carried over from a different problem or setting and 3 = represents that the passage is dedicated to the problem in the query and contains a method, formulation or finding that directly solves it. Important Instruction: Judge whether the passage could be used to solve the problem, not how similar its topic is: a passage from a different field can deserve category 2 or 3, and a passage on the same topic that does not help solve the problem deserves category 1. If none of the above satisfies give it category 0. Query: {query} Passage: {passage} For reference, here is a candidate analogical correspondence between two sets of objects and relations. Object mappings (left -> right): {mappings} Shared relations (left <-> right): {relations} Use it only if it applies: check whether the left-side objects and relations are present in the query's problem and the right-side ones in the passage. If they are not, ignore the correspondence. Split this problem into steps: Consider the underlying intent of the search. Measure how well the content matches a likely intent of the query (M). Measure how trustworthy the passage is (T). Consider the aspects above and the relative importance of each, and decide on a final score (O). Final score must be an integer value only. Do not provide any code in result. Provide each score in the format of: ##final score: score without providing any reasoning.
\end{promptbox}

\subsection{Usefulness Scoring for Hard Negatives}\label{app:prompt-hardneg}%
Hard-negative candidates are scored on the same 0--3 scale with the same prompt as in Appendix~\ref{app:prompt-usefulness}, except that no object mappings or shared relations are given.
We use GPT-6.1 Sol for this step.
The query is the generated query of $q$, and the passage is the title and abstract of the candidate (abstract truncated to 2,000 characters).
Only candidates with a score of 1 or lower are used as hard negatives; candidates scored 2 or higher are discarded.

\begin{promptbox}
Given a query and a passage, you must provide a score on an integer scale of 0 to 3 with the following meanings: 0 = represent that the passage has nothing to do with the query, 1 = represents that the passage seems related to the query but does not provide anything that could be used to solve its problem, 2 = represents that the passage has a method, formulation or finding that could be used to solve the problem in the query, but it is partial, unclear, or has to be carried over from a different problem or setting and 3 = represents that the passage is dedicated to the problem in the query and contains a method, formulation or finding that directly solves it. Important Instruction: Judge whether the passage could be used to solve the problem, not how similar its topic is: a passage from a different field can deserve category 2 or 3, and a passage on the same topic that does not help solve the problem deserves category 1. If none of the above satisfies give it category 0. Query: {query} Passage: {passage} Split this problem into steps: Consider the underlying intent of the search. Measure how well the content matches a likely intent of the query (M). Measure how trustworthy the passage is (T). Consider the aspects above and the relative importance of each, and decide on a final score (O). Final score must be an integer value only. Do not provide any code in result. Provide each score in the format of: ##final score: score without providing any reasoning.
\end{promptbox}

\section{Human Agreement}\label{app:human-agreement}
\begin{wraptable}{r}{0.48\linewidth}
\centering
\fontsize{8pt}{9pt}\selectfont
\setlength{\tabcolsep}{3pt}
\begin{tabularx}{\linewidth}{@{}Xccc@{}}
\toprule
Task & Items & Agreement & Cohen's $\kappa$ \\
\midrule
1. Transfer verification (P2) & 50 & 0.92 & 0.84 \\
2. Usefulness of positives (P3) & 25 & 0.92 & 0.84 \\
3. Usefulness of candidates (N2) & 25 & 0.96 & 0.92 \\
\bottomrule
\end{tabularx}
\caption{\textbf{Agreement between the LLM judges and a human annotator.} Usefulness scores are binarized at 2.}
\label{tab:human-agreement}
\end{wraptable}
We measure the agreement between the LLM judges and a human annotator, one of the authors, on two tasks of 50 items each.
\textbf{Task 1} covers the transfer verification of Step P2: pairs that passed the analogy judgment, half accepted and half rejected by the verification.
The annotator sees the titles and abstracts of $q$ and $d$, the sentences of $q$ citing $d$, and the extracted analogy, and answers yes or no.
\textbf{Task 2} covers usefulness scoring: 25 query--positive pairs scored by GPT-6 Astra in Step P3 and 25 query--candidate pairs scored by GPT-6.1 Sol in Step N2, each set split about evenly between scores of 2 or above and 1 or below (13 and 12 for P3, 12 and 13 for N2).
The annotator sees the query and the title and abstract of the paper, together with the extracted analogy when the pair has one, and gives a score from 0 to 3.
Each item has a different base paper (Task 1) or query (Task 2).
Scores are binarized at 2, and Table~\ref{tab:human-agreement} reports the agreement for each LLM judge.
The annotator followed the guidelines below.

\paragraph{Guidelines for Task 1.}
\begin{promptbox}
You see a problem paper q, a paper D that q cites, the sentences in q that cite D, and a first-pass judgment that already proposes object mappings, shared relations and a transfer path between them. Decide whether q carries over something from D to solve its own, different problem.
The first-pass judgment is a proposal to check, not an answer to follow: it is often wrong, and your job includes deciding whether the texts support it.
Yes (transfer) when all of these hold:
- q and D address different problems. D is not prior work on the same task as q.
- q actually uses, adapts or builds on a contribution of D (a method, formulation, component, measure, finding, design or resource) for its own problem, as shown by the texts.
- The proposed mappings that connect D's contribution to q's problem are supported by the texts.
- What is carried over serves what q is trying to achieve.
It does not matter whether the contribution is general-purpose or already common practice: a widely used technique still counts if it was carried over from D's different problem and helps q.
No (reject) if any of these holds:
- Same problem: D already addresses q's problem (same task, or the same problem line in another dataset, population, language or setting).
- Background only: q cites D only for motivation, related work, contrast or history, without using it.
- Unsupported: the proposed mappings or transfer are not supported by the texts.
Use only the texts shown. Do not rely on outside knowledge.
\end{promptbox}

\paragraph{Guidelines for Task 2.}
\begin{promptbox}
You see a query (a research problem) and a passage (a paper's title and abstract). Rate on 0-3 whether the passage could be used to solve the problem in the query:
- 0 = the passage has nothing to do with the query.
- 1 = the passage seems related to the query but does not provide anything that could be used to solve its problem.
- 2 = the passage has a method, formulation or finding that could be used to solve the problem in the query, but it is partial, unclear, or has to be carried over from a different problem or setting.
- 3 = the passage is dedicated to the problem in the query and contains a method, formulation or finding that directly solves it.
Important: judge whether the passage could be used to solve the problem, not how similar its topic is. A passage from a different field can deserve 2 or 3, and a passage on the same topic that does not help solve the problem deserves 1.
Some items also show a first-pass judgment: a candidate correspondence between two sets of objects and relations. Use it only if it applies -- check whether the left-side objects and relations are present in the query's problem and the right-side ones in the passage. If they are not, ignore it. Items without this block are judged on the query and passage alone.
\end{promptbox}

\section{OpenAlex}
\label{sec:appendix-openalex}

We use the OpenAlex snapshot of 2026-09-23.
Table~\ref{tab:openalex-examples} shows examples of the four levels of the OpenAlex hierarchy.

\begin{table}[t]
\centering
\small
\begin{tabularx}{\linewidth}{@{}l>{\raggedright\arraybackslash}X>{\raggedright\arraybackslash}X>{\raggedright\arraybackslash}X@{}}
\toprule
Domain & Field & Subfield & Topic \\
\midrule
Health Sciences & Dentistry & Periodontics & Oral microbiology and periodontitis research \\
 & Nursing & Nutrition and Dietetics & Child Nutrition and Water Access \\
\midrule
Life Sciences & Biochemistry, Genetics and Molecular Biology & Genetics & Genetic Associations and Epidemiology \\
 & Immunology and Microbiology & Immunology & Immunotherapy and Immune Responses \\
\midrule
Physical Sciences & Computer Science & Computer Vision and Pattern Recognition & Advanced Neural Network Applications \\
 & Mathematics & Statistics and Probability & Advanced Causal Inference Techniques \\
 & Chemical Engineering & Catalysis & Ammonia Synthesis and Nitrogen Reduction \\
\midrule
Social Sciences & Decision Sciences & Statistics, Probability and Uncertainty & Meta-analysis and systematic reviews \\
 & Economics, Econometrics and Finance & Economics and Econometrics & Market Dynamics and Volatility \\
 & Business, Management and Accounting & Strategy and Management & Innovation and Knowledge Management \\
\bottomrule
\end{tabularx}
\caption{\textbf{Examples of the OpenAlex hierarchy.} Each row is a topic assigned to papers in CPR-eval, with the subfield, field, and domain it belongs to. OpenAlex has 4 domains, 26 fields, 252 subfields, and about 4,500 topics.}
\label{tab:openalex-examples}
\end{table}

\begin{table}[t]
\centering
\begin{minipage}[t]{0.40\linewidth}
\centering
\small
\begin{tabular*}{\linewidth}{@{\extracolsep{\fill}}lrr}
\toprule
Level & Count & First divergence (\%) \\
\midrule
Domain & 4 & 45.3 \\
Field & 26 & 34.6 \\
Subfield & 233 & 12.4 \\
Topic & 2,724 & 7.7 \\
\bottomrule
\end{tabular*}
\caption{\textbf{Topical spread of CPR-Dataset pairs.} For each OpenAlex level, over both splits: the number of distinct units among base papers and positives (\textit{Count}), and the share of pairs whose base paper and positive first differ at that level (\textit{First divergence}).}
\label{tab:levels}
\end{minipage}\hfill
\begin{minipage}[t]{0.55\linewidth}
\centering
\small
\setlength{\tabcolsep}{2.5pt}
\begin{tabular*}{\linewidth}{@{\extracolsep{\fill}}lcrr}
\toprule
Dataset & \begin{tabular}[b]{@{}c@{}}Cross-\\topic\end{tabular} & Fields & Size \\
\midrule
SPECTER~\citep{cohan-etal-2020-specter} & $\times$ & 26 & 684K \\
SciDocs~\citep{cohan-etal-2020-specter} & $\times$ & 20 & 1,000 \\
LitSearch~\citep{ajith-etal-2024-litsearch} & $\times$ & 10 & 597 \\
ScholarCatalyst~\citep{kim2026scholarcatalystbenchmarkretrievingpapers} & $\times$ & 16 & 894 \\
MOOSE-Chem~\citep{yang2025moosechem} & $\times$ & 7 & 51 \\
ResearchBench~\citep{liu-etal-2026-researchbench} & $\times$ & 25 & 1,386 \\
SPDF~\citep{kulikowski2026problemdifferentfieldcrossdomain} & \checkmark & 14 & 92 \\
\midrule
\textbf{CPR-Dataset} & \checkmark & 26 & 11.9K \\
\bottomrule
\end{tabular*}
\caption{\textbf{Comparison between CPR-Dataset and existing scientific paper retrieval datasets.} \textit{Cross-topic} marks whether positives must come from a topic other than the query's. \textit{Fields} counts the OpenAlex fields of the papers that queries come from, after we reclassify these papers into OpenAlex fields. \textit{Size} counts evaluation queries, except for SPECTER, where it counts the citation-based triplets used to train the SPECTER model, and for CPR-Dataset, where it counts the 1,300 evaluation queries and 10,573 training pairs. SPDF denotes Same Problem, Different Field.}
\label{tab:comparison}
\end{minipage}
\end{table}
Table~\ref{tab:levels} shows how far the positives of CPR-Dataset lie from their base papers at each level of the OpenAlex hierarchy.

\section{Comparison with Existing Datasets}
\label{app:comparison}
Table~\ref{tab:comparison} compares CPR-Dataset with existing datasets for scientific paper retrieval.
Figure~\ref{fig:landscape-full} repeats Figure~\ref{fig:landscape} with every pair of each benchmark instead of the 100 with the longest lines: 4,741 pairs for SciDocs, 633 for LitSearch, 177 for Same Problem, Different Field, and 1,300 for CPR-eval.

\begin{figure}[t]
\centering
\includegraphics[width=\linewidth]{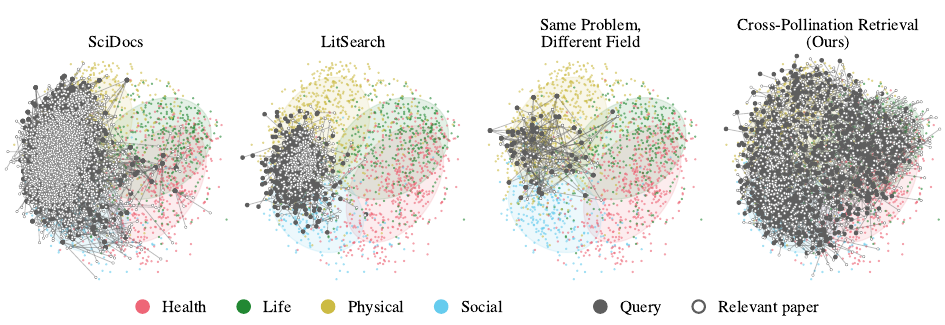}
\caption{\textbf{All query and relevant paper pairs of each benchmark.} The same map as Figure~\ref{fig:landscape}, with every pair instead of the 100 with the longest lines.}
\label{fig:landscape-full}
\end{figure}

\section{Metric and Loss Definitions}\label{sec:appendix-definitions}
\paragraph{Metrics.}
For a query, let $\mathrm{rank}(d)$ be the rank of a document $d$ in the retriever's ranking of the candidates, $d_i$ the document at rank $i$, and $G$ the set of relevant documents.
In CPR-eval, $G$ contains only the positive $d^{+}$, so
\begin{align*}
\mathrm{Recall@}k &= \mathbf{1}\!\left[\mathrm{rank}(d^{+}) \leq k\right],\\
\mathrm{MRR@10} &= \frac{\mathbf{1}\!\left[\mathrm{rank}(d^{+}) \leq 10\right]}{\mathrm{rank}(d^{+})}.
\end{align*}
BRIGHT and AnaloBench have several relevant documents per query, and we use nDCG@10 with binary gains:
\[
\mathrm{nDCG@10} = \frac{\sum_{i=1}^{10} \mathbf{1}[d_i \in G] / \log_2(i+1)}{\sum_{i=1}^{\min(|G|,10)} 1 / \log_2(i+1)}.
\]
Each metric is averaged over queries.
On BRIGHT, we average within each of the seven subsets and then over the subsets; on AnaloBench, we average within each story length and then over the three lengths.

\paragraph{Training loss.}
A training pair consists of a query $x$, its positive $d^{+}$, and its hard negatives $N$ (nine in CPR-train).
Let $s(x,d)$ be the cosine similarity between the embeddings of $x$ and $d$, and $B$ the positives and hard negatives of the other queries encoded together with $x$ (in-batch negatives): 31 other queries for Qwen3-Embedding (8 GPUs $\times$ 4) and 7 for llama-embed-nemotron-8b (8 GPUs $\times$ 1), since the batch of 128 queries is built by gradient accumulation (Table~\ref{tab:hparams}).
The InfoNCE loss~\citep{oord2019representationlearningcontrastivepredictive} of the pair is
\[
\mathcal{L}(x) = -\log \frac{\exp\left(s(x,d^{+})/T\right)}{\sum_{d \in \{d^{+}\} \cup N \cup B} \exp\left(s(x,d)/T\right)},
\]
averaged over the queries of a batch, with temperature $T = 0.1$ for Qwen3-Embedding and 0.02 for llama-embed-nemotron-8b (Table~\ref{tab:hparams}).
Training without hard negatives (Appendix~\ref{sec:appendix-no-hardneg}) uses the same loss with $N = \emptyset$.

\section{Full Evaluation Results}\label{sec:appendix-metrics}%
Table~\ref{tab:main} reports Recall@1 and Recall@10.
Table~\ref{tab:full-metrics} gives Recall@5 and MRR@10 in the same layout, and Table~\ref{tab:full-ci} gives 95\% confidence intervals for all four metrics over all queries. Table~\ref{tab:full-trained} gives the same four metrics for the retrievers trained on each training set. Tables~\ref{tab:full-bright} and~\ref{tab:full-analobench} give the results of the same trained retrievers on each subset of BRIGHT and AnaloBench.
The conclusions do not depend on the metric: on every metric over all queries, inf-retriever-v1-pro, Gemini Embedding, and LateOn are the three best retrievers, and the LLM reranker is far above all retrievers.

\paragraph{Input length.} The off-the-shelf dense retrievers with a length limit (Qwen3-Embedding-0.6B and 4B, SPECTER2, the OpenScholar retriever, and inf-retriever-v1-pro) encode at most 512 tokens in Tables~\ref{tab:main}, \ref{tab:full-metrics}, and~\ref{tab:full-ci}. The trained retrievers and their before-training baselines (Figures~\ref{fig:trained} and~\ref{fig:domain}, Tables~\ref{tab:full-trained} and~\ref{tab:heldout-r10}) encode at most 1,024 tokens, the length used in training (Appendix~\ref{sec:appendix-hparams}), so the before-training values differ slightly from those in Table~\ref{tab:main}.

\begin{table}[t]
\centering
\small
\setlength{\aboverulesep}{0pt}\setlength{\belowrulesep}{0pt}\renewcommand{\arraystretch}{1.2}
\begin{tabularx}{\linewidth}{l|YY|YY|YY|YY|YY}
\toprule
 & \multicolumn{8}{c|}{\textbf{Domain}} & \multicolumn{2}{c}{} \\
\cmidrule{2-9}
 & \multicolumn{2}{c|}{Health} & \multicolumn{2}{c|}{Life} & \multicolumn{2}{c|}{Physical} & \multicolumn{2}{c|}{Social} & \multicolumn{2}{c}{Avg.} \\
Method & R@5 & MRR & R@5 & MRR & R@5 & MRR & R@5 & MRR & R@5 & MRR \\
\midrule
\multicolumn{11}{c}{\textit{LLM reranker}$^{\dagger}$} \\
\midrule
MiMo-V2.6-Pro & 0.80 & 0.68 & 0.76 & 0.67 & 0.73 & 0.63 & 0.81 & 0.67 & 0.77 & 0.66 \\
\midrule
\multicolumn{11}{c}{\textit{Sparse}} \\
\midrule
BM25 & 0.09 & 0.06 & 0.12 & 0.07 & 0.12 & 0.07 & 0.12 & 0.10 & 0.11 & 0.07 \\
BGE-M3 (sparse) & 0.08 & 0.06 & 0.07 & 0.04 & 0.08 & 0.05 & 0.09 & 0.07 & 0.08 & 0.05 \\
\midrule
\multicolumn{11}{c}{\textit{Dense (general)}} \\
\midrule
Qwen3-Emb-0.6B & 0.09 & 0.07 & 0.07 & 0.05 & 0.07 & 0.05 & 0.09 & 0.07 & 0.08 & 0.06 \\
BGE-M3 (dense) & 0.08 & 0.06 & 0.06 & 0.04 & 0.06 & 0.04 & 0.08 & 0.06 & 0.07 & 0.05 \\
Qwen3-Emb-4B & 0.13 & 0.09 & 0.14 & 0.10 & 0.11 & 0.07 & 0.14 & 0.10 & 0.13 & 0.09 \\
Gemini Embedding & 0.17 & 0.13 & \textbf{0.22} & 0.16 & \textbf{0.18} & 0.13 & 0.20 & \textbf{0.16} & 0.19 & 0.14 \\
\midrule
\multicolumn{11}{c}{\textit{Dense (scientific)}} \\
\midrule
SPECTER2 & 0.05 & 0.04 & 0.06 & 0.05 & 0.06 & 0.04 & 0.03 & 0.02 & 0.05 & 0.04 \\
OpenScholar retriever & 0.09 & 0.05 & 0.08 & 0.05 & 0.08 & 0.06 & 0.10 & 0.07 & 0.09 & 0.06 \\
\midrule
\multicolumn{11}{c}{\textit{Dense (reasoning)}} \\
\midrule
inf-retriever-v1-pro & \textbf{0.21} & \textbf{0.17} & 0.21 & \textbf{0.17} & 0.17 & \textbf{0.14} & \textbf{0.22} & \textbf{0.16} & \textbf{0.20} & \textbf{0.15} \\
\midrule
\multicolumn{11}{c}{\textit{Late interaction}} \\
\midrule
LateOn & 0.16 & 0.12 & 0.14 & 0.10 & 0.15 & 0.11 & 0.17 & 0.12 & 0.15 & 0.12 \\
BGE-M3 (multi-vector) & 0.11 & 0.07 & 0.07 & 0.04 & 0.05 & 0.04 & 0.10 & 0.06 & 0.08 & 0.05 \\
\bottomrule
\end{tabularx}
\caption{\textbf{Recall@5 and MRR@10 of retrievers and an LLM reranker on CPR-eval.} Columns show results by the OpenAlex domain of the base paper and the average over all queries. The best retriever in each column is in bold. MRR denotes MRR@10. Random ranking gives 0.03 and 0.02. $^{\dagger}$Shown for reference: the LLM reranker reorders the same 151 candidates in a tournament.}
\label{tab:full-metrics}
\end{table}

\begin{table}[t]
\centering
\small
\setlength{\aboverulesep}{0pt}\setlength{\belowrulesep}{0pt}\renewcommand{\arraystretch}{1.2}
\begin{tabularx}{\linewidth}{l|YYYY}
\toprule
Method & R@1 & R@5 & R@10 & MRR@10 \\
\midrule
\multicolumn{5}{c}{\textit{LLM reranker}$^{\dagger}$} \\
\midrule
MiMo-V2.6-Pro & 0.58 {\scriptsize[0.55, 0.61]} & 0.77 {\scriptsize[0.74, 0.79]} & 0.81 {\scriptsize[0.79, 0.84]} & 0.66 {\scriptsize[0.63, 0.68]} \\
\midrule
\multicolumn{5}{c}{\textit{Sparse}} \\
\midrule
BM25 & 0.04 {\scriptsize[0.03, 0.05]} & 0.11 {\scriptsize[0.10, 0.13]} & 0.16 {\scriptsize[0.14, 0.19]} & 0.07 {\scriptsize[0.06, 0.09]} \\
BGE-M3 (sparse) & 0.03 {\scriptsize[0.02, 0.04]} & 0.08 {\scriptsize[0.06, 0.09]} & 0.11 {\scriptsize[0.09, 0.12]} & 0.05 {\scriptsize[0.04, 0.06]} \\
\midrule
\multicolumn{5}{c}{\textit{Dense (general)}} \\
\midrule
Qwen3-Emb-0.6B & 0.04 {\scriptsize[0.03, 0.05]} & 0.08 {\scriptsize[0.06, 0.09]} & 0.12 {\scriptsize[0.10, 0.13]} & 0.06 {\scriptsize[0.05, 0.07]} \\
BGE-M3 (dense) & 0.03 {\scriptsize[0.02, 0.04]} & 0.07 {\scriptsize[0.05, 0.08]} & 0.10 {\scriptsize[0.08, 0.11]} & 0.05 {\scriptsize[0.04, 0.06]} \\
Qwen3-Emb-4B & 0.06 {\scriptsize[0.04, 0.07]} & 0.13 {\scriptsize[0.11, 0.15]} & 0.17 {\scriptsize[0.15, 0.19]} & 0.09 {\scriptsize[0.07, 0.10]} \\
Gemini Embedding & 0.10 {\scriptsize[0.08, 0.12]} & 0.19 {\scriptsize[0.17, 0.21]} & 0.26 {\scriptsize[0.23, 0.28]} & 0.14 {\scriptsize[0.13, 0.16]} \\
\midrule
\multicolumn{5}{c}{\textit{Dense (scientific)}} \\
\midrule
SPECTER2 & 0.02 {\scriptsize[0.01, 0.02]} & 0.05 {\scriptsize[0.04, 0.07]} & 0.09 {\scriptsize[0.08, 0.11]} & 0.04 {\scriptsize[0.03, 0.04]} \\
OpenScholar retriever & 0.04 {\scriptsize[0.03, 0.05]} & 0.09 {\scriptsize[0.07, 0.10]} & 0.12 {\scriptsize[0.10, 0.14]} & 0.06 {\scriptsize[0.05, 0.07]} \\
\midrule
\multicolumn{5}{c}{\textit{Dense (reasoning)}} \\
\midrule
inf-retriever-v1-pro & 0.12 {\scriptsize[0.10, 0.13]} & 0.20 {\scriptsize[0.17, 0.22]} & 0.26 {\scriptsize[0.23, 0.28]} & 0.15 {\scriptsize[0.14, 0.17]} \\
\midrule
\multicolumn{5}{c}{\textit{Late interaction}} \\
\midrule
LateOn & 0.09 {\scriptsize[0.07, 0.10]} & 0.15 {\scriptsize[0.13, 0.17]} & 0.19 {\scriptsize[0.17, 0.21]} & 0.12 {\scriptsize[0.10, 0.13]} \\
BGE-M3 (multi-vector) & 0.03 {\scriptsize[0.03, 0.05]} & 0.08 {\scriptsize[0.06, 0.09]} & 0.11 {\scriptsize[0.09, 0.12]} & 0.05 {\scriptsize[0.04, 0.06]} \\
\bottomrule
\end{tabularx}
\caption{\textbf{Results on CPR-eval over all 1,300 queries with 95\% confidence intervals.} Brackets give the 2.5th and 97.5th percentiles over 2,000 bootstrap resamples of the queries. $^{\dagger}$Shown for reference: the LLM reranker reorders the same 151 candidates in a tournament.}
\label{tab:full-ci}
\end{table}

\begin{table}[t]
\centering
\small
\setlength{\aboverulesep}{0pt}\setlength{\belowrulesep}{0pt}\renewcommand{\arraystretch}{1.2}
\begin{tabularx}{\linewidth}{ll|YYYY}
\toprule
Model & Training data & R@1 & R@5 & R@10 & MRR@10 \\
\midrule
Qwen3-Emb-0.6B & Before training & 0.04 {\scriptsize[0.03, 0.05]} & 0.08 {\scriptsize[0.06, 0.09]} & 0.12 {\scriptsize[0.10, 0.14]} & 0.06 {\scriptsize[0.05, 0.07]} \\
 & CPR-train & \textbf{0.24} {\scriptsize[0.21, 0.26]} & \textbf{0.41} {\scriptsize[0.39, 0.44]} & \textbf{0.50} {\scriptsize[0.47, 0.53]} & \textbf{0.31} {\scriptsize[0.29, 0.33]} \\
 & MS MARCO & 0.05 {\scriptsize[0.03, 0.06]} & 0.09 {\scriptsize[0.08, 0.11]} & 0.12 {\scriptsize[0.11, 0.14]} & 0.07 {\scriptsize[0.05, 0.08]} \\
 & ReasonIR data & 0.03 {\scriptsize[0.02, 0.04]} & 0.07 {\scriptsize[0.06, 0.09]} & 0.11 {\scriptsize[0.10, 0.13]} & 0.05 {\scriptsize[0.04, 0.06]} \\
 & SPECTER data & 0.02 {\scriptsize[0.01, 0.03]} & 0.06 {\scriptsize[0.05, 0.07]} & 0.09 {\scriptsize[0.08, 0.11]} & 0.04 {\scriptsize[0.03, 0.05]} \\
\midrule
Qwen3-Emb-4B & Before training & 0.06 {\scriptsize[0.04, 0.07]} & 0.13 {\scriptsize[0.11, 0.15]} & 0.17 {\scriptsize[0.15, 0.19]} & 0.09 {\scriptsize[0.07, 0.10]} \\
 & CPR-train & \textbf{0.41} {\scriptsize[0.38, 0.43]} & \textbf{0.60} {\scriptsize[0.57, 0.62]} & \textbf{0.69} {\scriptsize[0.66, 0.71]} & \textbf{0.49} {\scriptsize[0.47, 0.52]} \\
 & MS MARCO & 0.06 {\scriptsize[0.05, 0.07]} & 0.11 {\scriptsize[0.10, 0.13]} & 0.16 {\scriptsize[0.14, 0.18]} & 0.08 {\scriptsize[0.07, 0.10]} \\
 & ReasonIR data & 0.06 {\scriptsize[0.05, 0.08]} & 0.13 {\scriptsize[0.12, 0.15]} & 0.18 {\scriptsize[0.16, 0.20]} & 0.09 {\scriptsize[0.08, 0.11]} \\
 & SPECTER data & 0.04 {\scriptsize[0.03, 0.05]} & 0.11 {\scriptsize[0.09, 0.13]} & 0.15 {\scriptsize[0.13, 0.17]} & 0.07 {\scriptsize[0.06, 0.08]} \\
\midrule
llama-embed-nemotron-8b & Before training & 0.09 {\scriptsize[0.07, 0.11]} & 0.18 {\scriptsize[0.16, 0.20]} & 0.25 {\scriptsize[0.23, 0.27]} & 0.13 {\scriptsize[0.12, 0.15]} \\
 & CPR-train & \textbf{0.43} {\scriptsize[0.40, 0.46]} & \textbf{0.62} {\scriptsize[0.60, 0.65]} & \textbf{0.70} {\scriptsize[0.68, 0.73]} & \textbf{0.51} {\scriptsize[0.49, 0.54]} \\
 & MS MARCO & 0.07 {\scriptsize[0.05, 0.08]} & 0.14 {\scriptsize[0.12, 0.15]} & 0.20 {\scriptsize[0.18, 0.23]} & 0.10 {\scriptsize[0.09, 0.11]} \\
 & ReasonIR data & 0.11 {\scriptsize[0.09, 0.13]} & 0.23 {\scriptsize[0.21, 0.26]} & 0.29 {\scriptsize[0.26, 0.31]} & 0.16 {\scriptsize[0.14, 0.18]} \\
 & SPECTER data & 0.05 {\scriptsize[0.04, 0.07]} & 0.15 {\scriptsize[0.13, 0.17]} & 0.20 {\scriptsize[0.18, 0.22]} & 0.09 {\scriptsize[0.08, 0.11]} \\
\bottomrule
\end{tabularx}
\caption{\textbf{Retrievers trained on each training set, evaluated on CPR-eval over all 1,300 queries.} Each model is trained for one epoch on 10,573 pairs with hard negatives and evaluated with a maximum length of 1,024 tokens. Brackets give 95\% bootstrap confidence intervals. Bold marks the best value for each model, including the model before training. For MS MARCO, ReasonIR data, and SPECTER data, the values are from one sample (seed 42); Table~\ref{tab:resample} gives the means over three samples.}
\label{tab:full-trained}
\end{table}

\begin{table}[t]
\centering
\small
\setlength{\aboverulesep}{0pt}\setlength{\belowrulesep}{0pt}\renewcommand{\arraystretch}{1.2}
\begin{tabularx}{\linewidth}{ll|YYYYYYY|Y}
\toprule
Model & Training data & Bio. & Earth & Econ. & Psy. & Rob. & Stack. & Sus. & Avg. \\
\midrule
Qwen3-Emb-0.6B & Before training & 0.113 & 0.211 & 0.158 & 0.136 & 0.111 & 0.120 & 0.116 & 0.138 \\
 & CPR-train & \textbf{0.233} & 0.222 & \textbf{0.215} & \textbf{0.248} & 0.094 & 0.177 & \textbf{0.141} & \textbf{0.190} \\
 & MS MARCO & 0.103 & \textbf{0.234} & 0.185 & 0.172 & \textbf{0.123} & 0.118 & 0.109 & 0.149 \\
 & ReasonIR data & 0.129 & 0.130 & 0.148 & 0.114 & 0.050 & 0.147 & 0.083 & 0.114 \\
 & SPECTER data & 0.131 & 0.192 & 0.183 & 0.170 & 0.098 & \textbf{0.187} & 0.107 & 0.153 \\
\midrule
Qwen3-Emb-4B & Before training & 0.125 & 0.246 & 0.136 & 0.181 & 0.109 & 0.148 & 0.111 & 0.151 \\
 & CPR-train & \textbf{0.357} & \textbf{0.336} & \textbf{0.265} & \textbf{0.361} & 0.077 & \textbf{0.227} & \textbf{0.172} & \textbf{0.256} \\
 & MS MARCO & 0.114 & 0.285 & 0.163 & 0.240 & 0.104 & 0.155 & 0.133 & 0.171 \\
 & ReasonIR data & 0.220 & 0.204 & 0.182 & 0.204 & 0.115 & 0.190 & 0.136 & 0.179 \\
 & SPECTER data & 0.175 & 0.244 & 0.173 & 0.197 & \textbf{0.142} & 0.190 & 0.139 & 0.180 \\
\midrule
llama-embed-nemotron-8b & Before training & 0.140 & 0.299 & 0.182 & 0.235 & 0.168 & 0.165 & 0.172 & 0.194 \\
 & CPR-train & \textbf{0.471} & \textbf{0.424} & \textbf{0.319} & \textbf{0.380} & 0.164 & \textbf{0.298} & 0.242 & \textbf{0.328} \\
 & MS MARCO & 0.153 & 0.328 & 0.227 & 0.303 & 0.171 & 0.205 & 0.167 & 0.222 \\
 & ReasonIR data & 0.179 & 0.293 & 0.182 & 0.270 & \textbf{0.239} & 0.263 & 0.231 & 0.237 \\
 & SPECTER data & 0.259 & 0.263 & 0.175 & 0.296 & 0.227 & 0.269 & \textbf{0.270} & 0.251 \\
\bottomrule
\end{tabularx}
\caption{\textbf{nDCG@10 on the seven StackExchange subsets of BRIGHT} (all 749 queries; each subset is searched over its own corpus). Bio.: biology, Earth: earth science, Econ.: economics, Psy.: psychology, Rob.: robotics, Stack.: Stack Overflow, Sus.: sustainable living. Avg. is the mean over the seven subsets. Bold marks the best value for each model, including the model before training. For MS MARCO, ReasonIR data, and SPECTER data, the values are from one sample (seed 42); Table~\ref{tab:resample} gives the means over three samples.}
\label{tab:full-bright}
\end{table}

\begin{table}[t]
\centering
\small
\setlength{\aboverulesep}{0pt}\setlength{\belowrulesep}{0pt}\renewcommand{\arraystretch}{1.2}
\begin{tabularx}{\linewidth}{ll|YYYY|YYYY}
\toprule
 & & \multicolumn{4}{c|}{T2 nDCG@10} & \multicolumn{4}{c}{T1 accuracy} \\
\cmidrule{3-10}
Model & Training data & S1 & S10 & S30 & Avg. & S1 & S10 & S30 & Avg. \\
\midrule
Qwen3-Emb-0.6B & Before training & 0.336 & 0.307 & 0.298 & 0.314 & 0.631 & 0.582 & 0.573 & 0.595 \\
 & CPR-train & \textbf{0.450} & \textbf{0.393} & \textbf{0.364} & \textbf{0.402} & \textbf{0.686} & \textbf{0.644} & \textbf{0.627} & \textbf{0.652} \\
 & MS MARCO & 0.369 & 0.264 & 0.259 & 0.297 & 0.653 & 0.541 & 0.546 & 0.580 \\
 & ReasonIR data & 0.265 & 0.257 & 0.264 & 0.262 & 0.553 & 0.502 & 0.522 & 0.526 \\
 & SPECTER data & 0.209 & 0.235 & 0.238 & 0.227 & 0.480 & 0.504 & 0.517 & 0.501 \\
\midrule
Qwen3-Emb-4B & Before training & 0.382 & 0.349 & 0.326 & 0.352 & 0.659 & 0.599 & 0.572 & 0.610 \\
 & CPR-train & \textbf{0.506} & \textbf{0.478} & \textbf{0.420} & \textbf{0.468} & \textbf{0.729} & \textbf{0.701} & \textbf{0.662} & \textbf{0.697} \\
 & MS MARCO & 0.367 & 0.264 & 0.258 & 0.296 & 0.663 & 0.554 & 0.552 & 0.590 \\
 & ReasonIR data & 0.332 & 0.297 & 0.262 & 0.297 & 0.593 & 0.548 & 0.531 & 0.557 \\
 & SPECTER data & 0.253 & 0.263 & 0.246 & 0.254 & 0.548 & 0.533 & 0.529 & 0.537 \\
\midrule
llama-embed-nemotron-8b & Before training & \textbf{0.514} & 0.374 & 0.362 & 0.417 & \textbf{0.702} & 0.616 & 0.622 & 0.647 \\
 & CPR-train & 0.513 & \textbf{0.433} & \textbf{0.421} & \textbf{0.455} & 0.693 & \textbf{0.650} & \textbf{0.634} & \textbf{0.659} \\
 & MS MARCO & 0.126 & 0.319 & 0.323 & 0.256 & 0.478 & 0.581 & 0.598 & 0.553 \\
 & ReasonIR data & 0.469 & 0.325 & 0.282 & 0.359 & 0.678 & 0.565 & 0.559 & 0.601 \\
 & SPECTER data & 0.449 & 0.343 & 0.332 & 0.375 & 0.663 & 0.582 & 0.585 & 0.610 \\
\bottomrule
\end{tabularx}
\caption{\textbf{Results on AnaloBench.} T2: retrieving analogous stories from a pool of stories (nDCG@10); T1: choosing the analogous story among four options (accuracy). S1, S10, and S30 are the story lengths in sentences; Avg. is their mean. Bold marks the best value for each model, including the model before training. For MS MARCO, ReasonIR data, and SPECTER data, the values are from one sample (seed 42); Table~\ref{tab:resample} gives the means over three samples.}
\label{tab:full-analobench}
\end{table}

\section{LLM Reranker Details}\label{sec:appendix-reranker}%
The LLM reranker in Table~\ref{tab:main} reorders the same 151 candidates of each query with MiMo-V2.6-Pro~\citep{mimo2026v26pro} in a tournament similar to those of TourRank~\citep{10.1145/3696410.3714863} and OBLIQ-Bench~\citep{OBLIQ-Bench}.
The candidates are shuffled with a fixed seed per query and shown with their title and abstract.
In the first round, the 151 candidates are split into eight groups of 18--19, and the model selects from each group, with a listwise prompt, the five candidates whose methods are most likely to solve the problem, leaving 40.
In the second round, these 40 are split into two groups of 20, and the model again selects five from each group, leaving ten finalists.
Each finalist is then scored separately from 0 to 10 with a pointwise prompt.
The finalists are ranked by score; when several finalists share the top score, one more listwise call orders them.
Candidates eliminated in the second round are ranked after the finalists, followed by those eliminated in the first round.
The two selection rounds use low reasoning effort, and the pointwise scoring and the tie-breaking call use the model's default.
Over the 1,300 queries, the reranker made 20.2 calls per query on average (usually 8 + 2 selection calls, 10 scoring calls, and one tie-breaking call when needed; ties at the top occurred for 362 queries), and the whole run cost \$91.78 (\$0.071 per query).
The positive reached the final ten for 1,058 queries and was eliminated in the second and first round for 134 and 108 queries, respectively. For 17 queries, a selection call could not be parsed, so only five finalists were scored (15 or 16 calls); in one of them, a positive eliminated in the second round still ranks eighth, which is why Recall@10 counts 1,059 queries.
The two prompts are shown below; \texttt{\{k\}} is 5 in the selection rounds and the number of tied finalists in the tie-breaking call.

\paragraph{Listwise selection.}
\begin{promptbox}
Below is a research problem and a list of candidate papers, each with an ID, title and abstract.
Rank the candidates by how likely the method or technique proposed in each paper could be used to solve the problem.

Research problem:
{query}

Candidates:
{candidates}

Before answering, briefly review the candidates: for the promising ones, note which part of the problem the paper's method could address and how.
Then end your answer with JSON giving the IDs of the {k} papers whose methods are most likely to solve the problem, best first:
{"top": [id, id, ...]}
\end{promptbox}

\paragraph{Pointwise scoring.}
\begin{promptbox}
Below is a research problem and one candidate paper with its title and abstract.
Judge how likely the method or technique proposed in the paper could be used to solve the problem.

Research problem:
{query}

Candidate paper:
{candidate}

Before answering, briefly note which part of the problem the paper's method could address and how, if any.
Then end your answer with JSON giving a score from 0 to 10, where 10 means the method could very likely be used to solve the problem and 0 means it could not:
{"score": n}
\end{promptbox}

\section{Training Hyperparameters}\label{sec:appendix-hparams}%
We fine-tune Qwen3-Embedding-0.6B and 4B~\citep{qwen3embedding} and llama-embed-nemotron-8b~\citep{babakhin2025llamaembednemotron8buniversaltextembedding} on CPR-train and on the three comparison training sets with the settings in Table~\ref{tab:hparams}.
For each model, all four training sets use the same settings; only the training data differ.
Qwen3-Embedding follows the official fine-tuning recipe of Qwen3-Embedding (ms-swift), and llama-embed-nemotron-8b follows the official NeMo AutoModel recipe, except for the maximum length (1,024 instead of 512), the number of hard negatives (nine instead of four), in-batch negatives across GPUs, the query prompt, and the seed.
There is no validation set: every model is trained for one epoch on all training pairs, and the final checkpoint is evaluated.
All models are trained on eight NVIDIA A100 80GB GPUs.
For BM25, we use the \texttt{bm25s} library~\citep{lu2024bm25s} with the Lucene variant of BM25 ($k_1=1.5$, $b=0.75$).

\begin{table}[t]
\centering
\small
\begin{tabularx}{\linewidth}{l>{\raggedright\arraybackslash}X>{\raggedright\arraybackslash}X}
\toprule
 & Qwen3-Embedding-0.6B / 4B & llama-embed-nemotron-8b \\
\midrule
Framework & ms-swift, full fine-tuning & NeMo AutoModel 0.8.0, full fine-tuning \\
Batch size & 128 queries (8 GPUs $\times$ 4 $\times$ 4 accumulation) & 128 queries (8 GPUs $\times$ 1 $\times$ 16 accumulation) \\
Steps & 83 (1 epoch) & 83 (1 epoch) \\
Learning rate & $6\times10^{-6}$, cosine, no warmup & $2\times10^{-6}$, 2 warmup steps, cosine to 1\% \\
Optimizer & AdamW ($\beta$ = 0.9, 0.95), weight decay 0.1 & AdamW ($\beta$ = 0.9, 0.999), weight decay 0.01 \\
Gradient clipping & 1.0 & 1.0 \\
Loss & InfoNCE, temperature 0.1 & InfoNCE, temperature 0.02 \\
Negatives & all hard negatives of the pair + in-batch & all hard negatives of the pair + in-batch (across 8 GPUs) \\
Max length & 1,024 tokens (query and document) & 1,024 tokens (query and document) \\
Pooling & last token, L2-normalized & mean, L2-normalized \\
Precision & bf16 & fp32 weights, bf16 computation \\
Query / document prompt & \texttt{Query:} / none & \texttt{Query:} / none \\
Seed & 42 & 42 \\
\bottomrule
\end{tabularx}
\caption{\textbf{Training hyperparameters.} The same settings are used for every training set. Evaluation uses the same maximum length (1,024 tokens) and query prompt, for the trained models and for the models before training.}
\label{tab:hparams}
\end{table}

\begin{wraptable}{r}{0.48\linewidth}
\centering
\small
\begin{tabularx}{\linewidth}{l>{\raggedright\arraybackslash}Xc}
\toprule
Training set & Source & Hard neg. \\
\midrule
CPR-train & ours & 9 \\
MS MARCO & \texttt{microsoft/ms\_marco} (v1.1, train) & 9 \\
ReasonIR data & \texttt{reasonir/reasonir-data} (HQ, VL) & 1 \\
SPECTER data & \texttt{allenai/scirepeval} (\texttt{cite\_prediction}) & 1 \\
\bottomrule
\end{tabularx}
\caption{\textbf{Training sets.} Each set has 10,573 pairs. \textit{Hard neg.} is the number of hard negatives per pair.}
\label{tab:hparams-data}
\end{wraptable}
\paragraph{Comparison training sets.}
Each comparison training set has the same number of pairs as CPR-train (10,573), sampled at random (seed 42) from the source in Table~\ref{tab:hparams-data}; Appendix~\ref{sec:appendix-resample} repeats the sampling with two more seeds.
MS MARCO~\citep{DBLP:journals/corr/NguyenRSGTMD16} lists a different number of retrieved passages for each query. To give its pairs the same form as CPR-train (one positive and nine hard negatives), we keep queries with exactly one selected passage and at least nine unselected passages, and use the first nine unselected passages (retrieved for the query but not selected by the annotators) as hard negatives.
ReasonIR and SPECTER provide one hard negative per pair, which we use as is.
For ReasonIR~\citep{shao2025reasonir}, we pool the HQ and VL subsets; the positives of HQ are BRIGHT document IDs, which we fill in with the BRIGHT documents.
For SPECTER, we use the citation triplets of the original SPECTER training data~\citep{cohan-etal-2020-specter} (query paper, cited paper, and one paper not cited by the query paper), as released in the \texttt{cite\_prediction} subset of SciRepEval~\citep{singh-etal-2023-scirepeval}, with each paper represented as ``title. abstract''.
In all sets, we keep only pairs whose query, positive, and negatives are non-empty, whose positive is not among its negatives, and whose negatives are distinct.
Because the number of hard negatives differs across training sets, we also train every model on the same pairs with all hard negatives removed (Appendix~\ref{sec:appendix-no-hardneg}), so that the training sets can also be compared with in-batch negatives only.

\section{Resampling the Comparison Training Sets}\label{sec:appendix-resample}
\begin{table}[t]
\centering
\small
\setlength{\tabcolsep}{3pt}
\begin{tabular}{lccc}
\toprule
Training set & CPR-eval & AnaloBench & BRIGHT \\
\midrule
\multicolumn{4}{l}{\textit{Qwen3-Embedding-0.6B}} \\
MS MARCO & .127$\,\pm\,$.004 & .297$\,\pm\,$.001 & .151$\,\pm\,$.006 \\
ReasonIR data & .119$\,\pm\,$.004 & .268$\,\pm\,$.010 & .119$\,\pm\,$.005 \\
SPECTER data & .092$\,\pm\,$.002 & .228$\,\pm\,$.001 & .154$\,\pm\,$.001 \\
\midrule
\multicolumn{4}{l}{\textit{Qwen3-Embedding-4B}} \\
MS MARCO & .162$\,\pm\,$.005 & .304$\,\pm\,$.008 & .180$\,\pm\,$.008 \\
ReasonIR data & .188$\,\pm\,$.010 & .299$\,\pm\,$.004 & .171$\,\pm\,$.007 \\
SPECTER data & .157$\,\pm\,$.003 & .264$\,\pm\,$.010 & .178$\,\pm\,$.005 \\
\midrule
\multicolumn{4}{l}{\textit{llama-embed-nemotron-8b}} \\
MS MARCO & .202$\,\pm\,$.005 & .261$\,\pm\,$.004 & .219$\,\pm\,$.003 \\
ReasonIR data & .288$\,\pm\,$.008 & .375$\,\pm\,$.015 & .221$\,\pm\,$.021 \\
SPECTER data & .207$\,\pm\,$.009 & .386$\,\pm\,$.009 & .247$\,\pm\,$.010 \\
\bottomrule
\end{tabular}
\caption{\textbf{Retrievers trained on three random samples of each comparison training set.} Mean $\pm$ sample standard deviation over seeds 42, 1, and 2, without leading zeros. CPR-eval: Recall@10 over all 1,300 queries. AnaloBench: T2 nDCG@10 averaged over S1, S10, and S30. BRIGHT: nDCG@10 averaged over the seven StackExchange subsets (all 749 queries).}
\label{tab:resample}
\end{table}
Each comparison training set is a random sample from its source, so its results may depend on the sample.
We therefore draw two more samples of 10,573 pairs from each source (seeds 1 and 2, in addition to seed 42) with the same procedure, and train and evaluate each model on them with the same settings.
Table~\ref{tab:resample} gives the mean and the sample standard deviation over the three samples, which Figure~\ref{fig:trained} shows.
The standard deviation is at most 0.021, below the smallest gap between CPR-train and an existing training set (0.036, on BRIGHT with Qwen3-Embedding-0.6B).
For MS MARCO, only 12,291 queries meet the conditions in Appendix~\ref{sec:appendix-hparams}, so its three samples share about 86\% of their pairs; the samples of ReasonIR and SPECTER share 1--3\%.

\section{Query Rewriting Details}\label{sec:appendix-rewriting}%
We apply three query rewriting methods with the prompts of their original papers.
All rewritings are generated by GLM-5.3~\citep{glm5team2026glm5vibecodingagentic}, an open-weight model, through OpenRouter with low reasoning effort and at most 4,000 output tokens including reasoning.
\textbf{HyDE}~\citep{gao-etal-2023-precise} generates eight hypothetical passages at temperature 0.7.
For BM25, the query and the eight passages are concatenated into one query; for Gemini Embedding, the query embedding and the embeddings of the eight passages are averaged.
\textbf{Query2doc}~\citep{wang-etal-2023-query2doc} generates one passage at temperature 1.0 with four in-context examples.
Each example pairs a query with the abstract of its positive; the four pairs come from our pilot data and do not appear in CPR-eval.
Following the original paper, the BM25 query is the query repeated five times followed by the passage; for Gemini Embedding, the query and the passage are joined with \texttt{[SEP]}.
\textbf{RAG-Fusion}~\citep{Rackauckas_2024} generates four additional queries at temperature 0.7.
The original query and the four generated queries are each used for retrieval, and the five rankings are combined with reciprocal rank fusion~\citep{cormack2009rrf} ($k = 60$).
All methods rank the same 151 candidates as the other retrievers.

\paragraph{Example.}
For the Psychology query ``How to capture the complex interrelationships between occupational stress and job satisfaction without including unnecessary connections'', the positive is a statistics paper, ``Extended Bayesian Information Criteria for Gaussian Graphical Models''.
BM25 ranks it 148th of 151 with the query alone, because the query shares almost no words with the paper.
The hypothetical passages generated by HyDE frame the problem as sparse network estimation and use the vocabulary of that field: five of the eight passages mention Gaussian graphical models, and six mention the extended Bayesian information criterion (EBIC).
For example, one passage reads: ``we propose employing sparse inverse covariance estimation, specifically the graphical lasso (glasso) procedure, which applies an L1 penalty to the inverse covariance matrix \ldots{} the resulting Gaussian graphical model captures the essential structure of the stress--satisfaction system.''
With these passages, BM25 ranks the positive first.
Here, the generator itself supplies the cross-field connection that the retriever would otherwise have to make.

\paragraph{HyDE.}
\begin{promptbox}
Please write a scientific paper passage to answer the question
Question: {query}
Passage:
\end{promptbox}

\paragraph{Query2doc.}
\begin{promptbox}
Write a passage that answers the given query:

Query: How to perform sentiment sequence tagging accurately while requiring less supervision?
Passage: In the primate visual system, the identification of objects and the processing of spatial information are accomplished by different cortical pathways. The computational properties of this two-systems design were explored by constructing simplifying connectionist models. The models were designed to simultaneously classify and locate shapes that could appear in multiple positions in a matrix, and the ease of forming representations of the two kinds of information was measured. Some networks were designed so that all hidden nodes projected to all output nodes, whereas others had the hidden nodes split into two groups, with some projecting to the output nodes that registered shape identity and the remainder projecting to the output nodes that registered location. The simulations revealed that splitting processing into separate streams for identifying and locating a shape led to better performance only under some circumstances. Provided that enough computational resources were available in both streams, split networks were able to develop more efficient internal representations, as revealed by detailed analyses of the patterns of weights between connections.

Query: How to reliably determine which characteristics of analysts' questions during earnings calls genuinely relate to those analysts' prior stock recommendations, without mistaking chance associations for real ones?
Passage: Mental illnesses, such as depression and post traumatic stress disorder (PTSD), are highly underdiagnosed globally. Populations sharing similar demographics and personality traits are known to be more at risk than others. In this study, we characterise the language use of users disclosing their mental illness on Twitter. Language-derived personality and demographic estimates show surprisingly strong performance in distinguishing users that tweet a diagnosis of depression or PTSD from random controls, reaching an area under the receiveroperating characteristic curve ‐ AUC ‐ of around .8 in all our binary classification tasks. In fact, when distinguishing users disclosing depression from those disclosing PTSD, the single feature of estimated age shows nearly as strong performance (AUC = .806) as using thousands of topics (AUC = .819) or tens of thousands of n-grams (AUC = .812). We also find that differential language analyses, controlled for demographics, recover many symptoms associated with the mental illnesses in the clinical literature.

Query: How to compose a sentence's word vectors into a single representation that captures how the words relate to one another?
Passage: In this work, we explore finite-dimensional linear representations of nonlinear dynamical systems by restricting the Koopman operator to an invariant subspace spanned by specially chosen observable functions. The Koopman operator is an infinite-dimensional linear operator that evolves functions of the state of a dynamical system. Dominant terms in the Koopman expansion are typically computed using dynamic mode decomposition (DMD). DMD uses linear measurements of the state variables, and it has recently been shown that this may be too restrictive for nonlinear systems. Choosing the right nonlinear observable functions to form an invariant subspace where it is possible to obtain linear reduced-order models, especially those that are useful for control, is an open challenge. Here, we investigate the choice of observable functions for Koopman analysis that enable the use of optimal linear control techniques on nonlinear problems. First, to include a cost on the state of the system, as in linear quadratic regulator (LQR) control, it is helpful to include these states in the observable subspace, as in DMD. However, we find that this is only possible when there is a single isolated fixed point, as systems with multiple fixed points or more complicated attractors are not globally topologically conjugate to a finite-dimensional linear system, and cannot be represented by a finite-dimensional linear Koopman subspace that includes the state. We then present a data-driven strategy to identify relevant observable functions for Koopman analysis by leveraging a new algorithm to determine relevant terms in a dynamical system by ℓ1-regularized regression of the data in a nonlinear function space; we also show how this algorithm is related to DMD. Finally, we demonstrate the usefulness of nonlinear observable subspaces in the design of Koopman operator optimal control laws for fully nonlinear systems using techniques from linear optimal control.

Query: How to determine how a question's wording affects the number of answers it receives, independent of when the question was asked.
Passage: Video dissemination through sites such as YouTube can have widespread impacts on opinions, thoughts, and cultures. Not all videos will reach the same popularity and have the same impact. Popularity differences arise not only because of differences in video content, but also because of other "content-agnostic" factors. The latter factors are of considerable interest but it has been difficult to accurately study them. For example, videos uploaded by users with large social networks may tend to be more popular because they tend to have more interesting content, not because social network size has a substantial direct impact on popularity. In this paper, we develop and apply a methodology that is able to accurately assess, both qualitatively and quantitatively, the impacts of various content-agnostic factors on video popularity. When controlling for video content, we observe a strong linear "rich-get-richer" behavior, with the total number of previous views as the most important factor except for very young videos. The second most important factor is found to be video age. We analyze a number of phenomena that may contribute to rich-get-richer, including the first-mover advantage, and search bias towards popular videos. For young videos we find that factors other than the total number of previous views, such as uploader characteristics and number of keywords, become relatively more important. Our findings also confirm that inaccurate conclusions can be reached when not controlling for content.

Query: {query}
Passage:
\end{promptbox}

\paragraph{RAG-Fusion.}
\begin{promptbox}
You are a helpful assistant that generates multiple search queries based on a single input query.
Generate multiple search queries related to: {query}
OUTPUT (4 queries):
\end{promptbox}

\section{Held-out Domain Setup}\label{sec:appendix-heldout}%
\begin{figure}[t]
\centering
\begin{minipage}[b]{0.48\linewidth}
\centering
\small
\begin{tabular*}{\linewidth}{@{\extracolsep{\fill}}lrrrr}
\toprule
Held-out & Query & Positive & Remaining & Used \\
\midrule
Health & 2,137 & 1,126 & 7,240 & 5,896 \\
Life & 2,963 & 1,217 & 6,323 & 5,896 \\
Physical & 2,957 & 1,650 & 5,896 & 5,896 \\
Social & 2,446 & 692 & 7,365 & 5,896 \\
\bottomrule
\end{tabular*}
\captionof{table}{\textbf{Training pairs per held-out domain.} Starting from 10,503 pairs, \textit{Query} pairs are removed because the query is in the held-out domain and \textit{Positive} pairs because only the positive is. \textit{Used} is a random sample of the remaining pairs of the size of the smallest fold.}
\label{tab:heldout-folds}
\end{minipage}\hfill
\begin{minipage}[b]{0.48\linewidth}
\centering
\includegraphics[width=\linewidth]{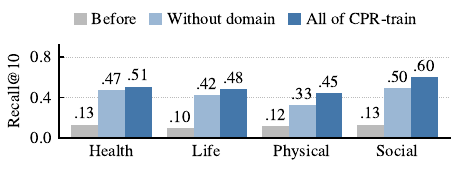}
\captionof{figure}{\textbf{Recall@10 of Qwen3-Embedding-0.6B on domains unseen in training.} The same setting as Figure~\ref{fig:domain}.}
\label{fig:domain06}
\end{minipage}
\end{figure}
For each of the four OpenAlex domains, we train a retriever without that domain and evaluate it on the queries of that domain.
The domain of a query is the domain of its base paper's field, and the domain of a positive is the domain of its primary OpenAlex topic.
We drop the 70 training pairs whose positive could not be matched to an OpenAlex work by DOI, PubMed, or MAG identifier, leaving 10,503 pairs.
To prevent leakage, a training pair is removed when its query or its positive belongs to the held-out domain; the domains of hard negatives are not checked.
We also confirm that no training query or positive is a query or positive of the evaluation set.
Because the folds differ in size after this removal, we randomly sample 5,896 pairs per fold (the size of the smallest fold, seed 42) so that all folds use the same amount of training data (Table~\ref{tab:heldout-folds}).
Each fold is trained for one epoch with the same settings as the main training runs (Appendix~\ref{sec:appendix-hparams}), and the final checkpoint is evaluated.
Evaluation uses the CPR-eval queries of the held-out domain with the same 151 candidates.
Table~\ref{tab:heldout-r10} lists Recall@10 with 95\% confidence intervals.

\begin{table}[t]
\centering
\small
\begin{tabular}{llcccc}
\toprule
Model & Training data & Health (250) & Life (250) & Physical (500) & Social (300) \\
\midrule
Qwen3-Emb-0.6B & Before training & 0.13 {\scriptsize[0.09, 0.18]} & 0.10 {\scriptsize[0.06, 0.14]} & 0.12 {\scriptsize[0.09, 0.14]} & 0.13 {\scriptsize[0.10, 0.17]} \\
 & Without the domain & 0.47 {\scriptsize[0.41, 0.54]} & 0.42 {\scriptsize[0.36, 0.48]} & 0.33 {\scriptsize[0.28, 0.37]} & 0.50 {\scriptsize[0.44, 0.56]} \\
 & All domains & 0.51 {\scriptsize[0.45, 0.57]} & 0.48 {\scriptsize[0.42, 0.54]} & 0.45 {\scriptsize[0.40, 0.49]} & 0.60 {\scriptsize[0.54, 0.66]} \\
\midrule
Qwen3-Emb-4B & Before training & 0.18 {\scriptsize[0.13, 0.22]} & 0.16 {\scriptsize[0.12, 0.21]} & 0.16 {\scriptsize[0.13, 0.19]} & 0.18 {\scriptsize[0.14, 0.22]} \\
 & Without the domain & 0.71 {\scriptsize[0.65, 0.76]} & 0.62 {\scriptsize[0.56, 0.67]} & 0.54 {\scriptsize[0.49, 0.58]} & 0.72 {\scriptsize[0.67, 0.77]} \\
 & All domains & 0.68 {\scriptsize[0.63, 0.74]} & 0.67 {\scriptsize[0.61, 0.73]} & 0.66 {\scriptsize[0.62, 0.70]} & 0.75 {\scriptsize[0.70, 0.80]} \\
\bottomrule
\end{tabular}
\caption{\textbf{Recall@10 on each held-out domain.} \textit{Without the domain}: trained on 5,896 pairs without the evaluated domain. \textit{All domains}: trained on all 10,573 training pairs, including the evaluated domain. Brackets give 95\% bootstrap confidence intervals; the number of queries is shown next to each domain.}
\label{tab:heldout-r10}
\end{table}

\section{Training without Hard Negatives}\label{sec:appendix-no-hardneg}%
\label{sec:appendix-inbatch}
To isolate the contribution of the hard negatives in CPR-train, we train every model on the same pairs of each training set with all hard negatives removed, so that only in-batch negatives are used; all other settings are as in Appendix~\ref{sec:appendix-hparams}.
Figure~\ref{fig:inbatch} compares these retrievers, and Table~\ref{tab:inbatch} lists the values next to those of the retrievers trained with hard negatives.

\begin{table}[t]
\centering
\small
\setlength{\aboverulesep}{0pt}\setlength{\belowrulesep}{0pt}\renewcommand{\arraystretch}{1.2}
\begin{tabularx}{\linewidth}{ll|YY|YY|YY}
\toprule
 & & \multicolumn{2}{c|}{\makecell{CPR-eval\\(R@10)}} & \multicolumn{2}{c|}{\makecell{AnaloBench\\(nDCG@10)}} & \multicolumn{2}{c}{\makecell{BRIGHT\\(nDCG@10)}} \\
Model & Training data & HN & IB & HN & IB & HN & IB \\
\midrule
Qwen3-Emb-0.6B & Before training & \multicolumn{2}{c|}{0.120} & \multicolumn{2}{c|}{0.314} & \multicolumn{2}{c}{0.138} \\
 & CPR-train & 0.501 & 0.323 & 0.402 & 0.330 & 0.190 & 0.176 \\
 & MS MARCO & 0.124 & 0.093 & 0.297 & 0.224 & 0.149 & 0.131 \\
 & ReasonIR data & 0.115 & 0.115 & 0.262 & 0.249 & 0.114 & 0.082 \\
 & SPECTER data & 0.094 & 0.094 & 0.227 & 0.224 & 0.153 & 0.155 \\
\midrule
Qwen3-Emb-4B & Before training & \multicolumn{2}{c|}{0.167} & \multicolumn{2}{c|}{0.352} & \multicolumn{2}{c}{0.151} \\
 & CPR-train & 0.689 & 0.487 & 0.468 & 0.376 & 0.256 & 0.212 \\
 & MS MARCO & 0.158 & 0.117 & 0.296 & 0.221 & 0.171 & 0.135 \\
 & ReasonIR data & 0.181 & 0.171 & 0.297 & 0.297 & 0.179 & 0.156 \\
 & SPECTER data & 0.153 & 0.150 & 0.254 & 0.248 & 0.180 & 0.180 \\
\midrule
llama-embed-nemotron-8b & Before training & \multicolumn{2}{c|}{0.249} & \multicolumn{2}{c|}{0.417} & \multicolumn{2}{c}{0.194} \\
 & CPR-train & 0.700 & 0.444 & 0.455 & 0.441 & 0.328 & 0.293 \\
 & MS MARCO & 0.203 & 0.218 & 0.256 & 0.419 & 0.222 & 0.211 \\
 & ReasonIR data & 0.288 & 0.272 & 0.359 & 0.379 & 0.237 & 0.222 \\
 & SPECTER data & 0.198 & 0.195 & 0.375 & 0.373 & 0.251 & 0.232 \\
\bottomrule
\end{tabularx}
\caption{\textbf{Retrievers trained with hard negatives (HN) and with in-batch negatives only (IB).} CPR-eval: Recall@10 over all 1,300 queries. AnaloBench: T2 nDCG@10 averaged over S1, S10, and S30. BRIGHT: nDCG@10 averaged over the seven StackExchange subsets (all 749 queries). The same pairs are used in both settings; IB removes their hard negatives. For MS MARCO, ReasonIR data, and SPECTER data, the values are from one sample (seed 42).}
\label{tab:inbatch}
\end{table}

\begin{figure}[t]
\centering
\includegraphics{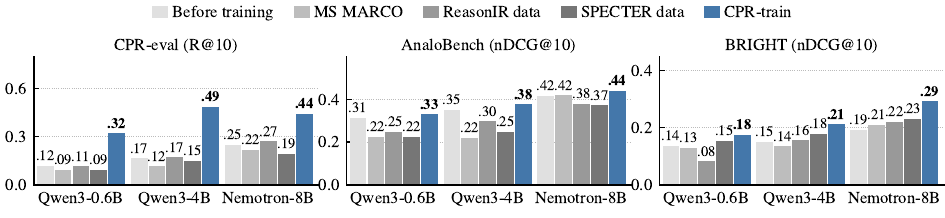}
\caption{\textbf{Retrievers trained with in-batch negatives only.} The same comparison as Figure~\ref{fig:trained}, but each retriever is trained without hard negatives. Before-training values are the same as in Figure~\ref{fig:trained}.}
\label{fig:inbatch}
\end{figure}

\section{Qualitative Examples}
\label{app:qualitative}
We compare five versions of Qwen3-Embedding-4B: before training, and after training on CPR-train, MS MARCO, ReasonIR data, or SPECTER data (one sample of each, seed 42, as in Table~\ref{tab:full-trained}). A retriever solves a query when it ranks a relevant document in its top 10 on BRIGHT or in its top 5 on AnaloBench. Only the retriever trained on CPR-train solves 75 of the 749 BRIGHT queries, and 12, 15, and 20 AnaloBench queries for stories of 1, 10, and 30 sentences. Table~\ref{tab:qualitative} shows five of these queries, chosen among those where this retriever ranks a relevant document first or second and none of the other four ranks one in its top 10.
On BRIGHT, the other retrievers rank first documents that share words with the query, whereas the relevant document uses none of these words and explains the underlying mechanism.
For instance, the question whether wind is nitrogen particles moving in one direction needs the idea of a net drift on top of random thermal motion, which the article on Brownian motion gives and the article on wind does not.
On AnaloBench, the query and its relevant stories share only their relational structure, not their setting.

\begin{table}[t]
\centering
\small
\renewcommand{\arraystretch}{1.15}
\begin{tabularx}{\linewidth}{@{}>{\raggedright\arraybackslash}p{0.21\linewidth}>{\raggedright\arraybackslash}X>{\raggedright\arraybackslash}X>{\raggedright\arraybackslash}X r@{}}
\toprule
Query (shortened) & Relevant document & Idea that links them & Ranked first by the other retrievers & Rank \\
\midrule
\multicolumn{5}{@{}l}{\textit{BRIGHT, earth science}} \\
Is wind just mainly nitrogen particles moving in one direction? & Wikipedia article on Brownian motion & a net drift on top of random thermal motion & Wikipedia article on wind; a page on air molecules hitting a moving car & 1 / 636 \\
\addlinespace[0.5em]
Why were temperature anomalies negative between 1950 and 1980, although oil consumption had been massive since 1880? & Wikipedia article on global dimming & aerosols from air pollution blocked sunlight & Pages on temperature anomalies and on climate change & 2 / 220 \\
\midrule
\multicolumn{5}{@{}l}{\textit{BRIGHT, robotics}} \\
In ROS2, do I need to rebuild the package after each change to a Python launch file? & A forum answer on symlink-install & linking the installed files to the source, so that edits take effect without compiling again & A page on the \texttt{ros2 launch} command; a tutorial that reruns colcon after adding a launch file & 1 / >1000 \\
\midrule
\multicolumn{5}{@{}l}{%
  \shortstack[l]{%
    \textit{AnaloBench,}\\
    \textit{stories of 10 sentences}%
  }%
} \\
Two parents say they are not good parents, yet never miss any of their children's events & A quiet student who is underrated ranks second in her class, among other stories & what people do shows the truth that their words deny & A practical husband tells his excited wife that a sudden work project means they must move their Hawaii vacation, and she accepts the change & 1 / 14 \\
\addlinespace[0.5em]
A student with nothing left to do during a break keeps thinking about his former girlfriend & A girl torments her grandmother on afternoons with nothing to do, among other stories & idle time leads to trouble & A woman who expects her boyfriend of several years to propose daydreams about a Paris honeymoon and wants to book it before he has even asked & 1 / 11 \\
\bottomrule
\end{tabularx}
\caption{\textbf{Queries solved only by the retriever trained on CPR-train.} All retrievers are Qwen3-Embedding-4B. \textit{Rank}: rank of the best relevant document for the retriever trained on CPR-train / the best of the same retriever trained on MS MARCO, ReasonIR data, or SPECTER data (one sample each, seed 42) or before training. BRIGHT ranks the documents of its subset; ``>1000'' means that the relevant document is not in the top 1,000. AnaloBench ranks a pool of 200 stories.}
\label{tab:qualitative}
\end{table}

}

\end{document}